\documentclass[10pt,aps,prl,twocolumn,superscriptaddress,showkeys]{revtex4}

\usepackage{graphicx}
\usepackage{amsmath, amsthm, amssymb}
\usepackage{epsfig}
\usepackage{dcolumn}
\usepackage{rotating}
\usepackage{multirow}
\usepackage{epsfig}
\usepackage{color}
\usepackage{float}
\usepackage{setspace}

\begin{document}

\title{Tomography of scaling}

\author{Marc Barthelemy}
\affiliation{Institut de Physique Th\'eorique, CEA-CNRS, F-91191 Gif-sur-Yvette, France}
\affiliation{CAMS (CNRS/EHESS) 54, boulevard Raspail, F-75006 Paris, France\\
\texttt{\small marc.barthelemy@ipht.fr}}

\date{\today}                                           

\begin{abstract}

  Scaling describes how a given quantity $Y$ that characterizes a system varies with its size $P$. For most complex systems it is of the form $Y\sim P^\beta$ with a nontrivial value of the exponent $\beta$, usually determined by regression methods. The presence of noise can make it difficult to conclude about the existence of a non-linear behavior with $\beta\neq 1$ and we propose here to circumvent fitting problems by investigating how two different systems of sizes $P_1$ and $P_2$ are related to each other. This leads us to define a local scaling exponent $\beta_{\mathrm{loc}}$ that we study versus the ratio $P_2/P_1$ and provides some sort of `tomography scan' of scaling across different values of the size ratio, allowing us to assess the relevance of nonlinearity in the system and to identify an effective exponent that minimizes the error for predicting the value of $Y$. We illustrate this method on various real-world datasets for cities and show that our method reinforces in some cases the standard analysis, but is also able to provide new insights in inconclusive cases and to detect problems in the scaling form such as the absence of a single scaling exponent or the presence of threshold effects.

\end{abstract}


\maketitle


\section{Introduction}

Scaling laws and associated scaling exponents are fundamental objects. Used in biology in order to understand how the metabolic rate varies with body size \cite{Kleiber:1932,Kleiber:1947}, scaling was widely used in physics to understand polymers \cite{degennes:1979}, phase transitions \cite{Goldenfeld:1992}, fluid dynamics and turbulence \cite{Barenblatt:2003}. Scaling also became a central tool for describing macroscopic properties of complex systems \cite{West:1997,Pumain:2004,Bettencourt:2007a} for mainly two reasons: First, the existence of a scaling law points to self-similarity: the system reproduces itself as the scale change. Second, these exponents constitute also precious guides for identifying critical factors and mechanisms in complex systems. In particular, when they cannot be deduced from simple dimensional considerations, they point to relevant scales and ingredients.

Given the simplicity of scaling measures, it is tempting to use this approach in order to get a first grasp about the behavior of complex systems that in general comprise a large number of constituents that interact with each other over various spatial and temporal scales. This is particularly true for urban systems for which we have now an abundance of data but are still lacking quantitative models for many aspects \cite{Batty:2013,Barthelemy:2016,Barthelemy:2019}. For cities, the scaling problem is to understand how extensive quantities vary with the size of the city, usually measured by its population \cite{Pumain:2004,Bettencourt:2007a}. Although our theoretical discussion is very general and could in principle be applied to any system we will use here the language of cities and apply our method to urban data. We thus consider a macroscopic quantity $Y$ that describes a given aspect of cities which can be socio-economical, about infrastructures, etc. and ask how it varies with the population $P$ of the city (according to a given definition of the city, see \cite{Arcaute:2015} and below for a discussion about this point). Empirical results for various quantities in cities were compiled for the first time in \cite{Bettencourt:2007a} and provided evidence that many quantities follow the scaling relation 
\begin{align}
  Y=aP^\beta
  \label{eq:pl}
\end{align} 
where $a$ is a prefactor and where the exponent $\beta$ is in general positive. This relation implies that the quantity per capita behaves as $Y/P\sim P^{\beta-1}$ and in the linear case ($\beta=1$) the quantity per capita is independent from the size of the city. This is in contrast with all the other cases ($\beta\neq 1$) where $Y/P$ depends on $P$ which means that there is a (nonlinear) effect of interactions in the city. It is therefore crucial to distinguish the case $\beta=1$ from $\beta\neq 1$ as it will determine how we model and understand the city. In the seminal paper \cite{Bettencourt:2007a}, it was shown that we have three different classes of quantities according to the value of $\beta$ and that correspond to different processes. As we just noted $\beta=1$ is the linear case for which the size of city has no impact - think of human related quantities for example - while for $\beta<1$ we mostly have infrastructure quantities denoting an economy of scale and for $\beta>1$ a positive effect of interactions (as expected for creative processes, social interaction dependent quantities such as innovations, or unfortunately negative aspects such as crimes or epidemic spreading). This study triggered a very large number of subsequent works that are difficult to cite all here, but we can mention scaling for roads properties \cite{Samaniego:2008},  for green space areas \cite{Fuller:2009}, for urban supply networks \cite{Kuhnert:2006}, for CO$_2$ emissions in cities \cite{Fragkias:2013,Rybski:2013,Oliveira:2014,Oliveira:2019}, for interaction activity \cite{Rybski:2009}, wealth, innovation and crimes
\cite{Bettencourt:2007b,Bettencourt:2010,Lobo:2013,Alves:2013,Nomaler:2014,Strano:2016,Caminha:2017}, etc. These different results motivated the search for a theoretical understanding and modelling that can explain these values \cite{Pumain:2006,Bettencourt:2013,Bettencourt:2013b,Louf:2013,Louf:2014,Verbavatz:2019,Molinero:2019}.

\section{Problems with fitting}

\begin{figure}[ht!]
\includegraphics[width=0.9\linewidth]{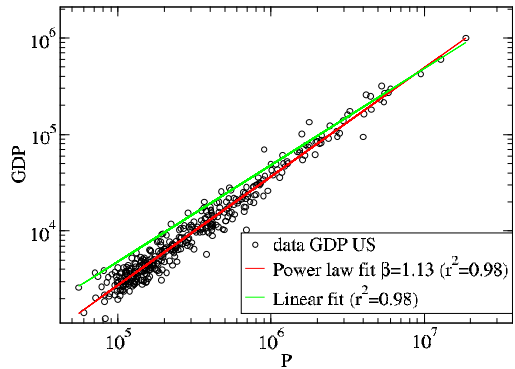}
\caption{ GDP (in millions of current dollars) in 2010 for Metropolitan Statistical Areas versus their population. We show here both the linear and the nonlinear fits. At this point it is difficult to conclude about a possible nonlinear behavior. Data from the Bureau of Economic Analysis \cite{bea}.}
\label{fig:gdpfits}
\end{figure}

The usual (and simplest) way to determine an estimator $\hat{\beta}$ of the scaling exponent $\beta$ is the ordinary least square regression, which consists essentially in plotting $Y$ versus $P$ in loglog and to find a power law fit (that is linear in loglog) such that the error (measured as the sum of squared differences) is minimum.  This is the classical method used throughout many different fields and poses no problem if (i) there are enough decades on both axes, (ii) there is not too much noise. If we are interested in the existence of nonlinearity (which is the case for cities), we can also add the constraint (iii) that  the exponent should be clearly different from one. These conditions are however unfortunately not always met. As an example we plot the GDP for each city in the US (for the year 2010) versus population and the result is shown in Fig.~\ref{fig:gdpfits}.
We basically have two decades of variation on both axis (which, roughly speaking, is the minimum in order to determine a power law exponent) and a reasonable amount of noise, leading to a power law fit that gives $\hat{\beta}\approx 1.13$ (with $r^2=0.98$). We see here that conditions (i) and (iii) are not met and we can only place a relative confidence in this value $1.13$. Indeed, a linear fit, which has one parameter less compared to the nonlinear fit, is also good (Fig.~\ref{fig:gdpfits}). More generally, involved statistical methods need then to be invoked if we can reject or not the linear assumption and this was the point of the excellent paper \cite{Leitao:2016}. These authors tested the hypothesis that observations are compatible with a nonlinear behavior and their conclusion for various quantities is that the estimate of $\beta$ together with confidence intervals depend a lot on fluctuations in the data and how they are modelled. It is thus difficult to get a clear-cut answer to the fundamental question if $\beta$ is different from one or not. These fitting problems were also discussed in \cite{Shalizi:2011} on GDP and income in the US where it was argued that other scaling forms could be used and that non-trivial scaling exponent values could be an artifact of using extensive quantities instead of intensive ones (per-capita rates). 

These problems were reinforced by other studies \cite{Oliveira:2014,Arcaute:2015,Cottineau:2017} that showed the importance of the definition of cities: these authors developed a framework for defining cities using commuters number and population density thresholds and could show on a UK dataset that many urban indicators scale linearly with population size, independently from the definition of urban boundaries. For quantities that display a nonlinear behavior, the scaling exponent value fluctuates considerably, and more importantly can be either larger or less than one according to the definition used (a problem also observed on the case of CO$_2$ emissions by transport \cite{Louf:2014}). In addition to these empirical problems, we also mention a study on congestion induced delays in cities which seem to scale with an exponent that varies in time, posing in fine the problem of mixing different cities at different stages of their evolution \cite{Depersin:2018}.

From an Ockham's razor perspective \cite{Ockham} choosing between a linear behavior independent from urban boundaries or a nonlinear scaling exponent whose value fluctuates considerably, leads to the conclusion that many socio-economical indicators are described by a linear behavior with $\beta=1$. This is however not a scientific proof and as our capacity for understanding cities relies crucially on this exponent value, the question is still somehow open and begs for a more satisfying answer. As most statistical frameworks and approaches lead to conclusions that depend critically on assumptions, especially in `gray cases' (with large noise, few decades for fitting, exponent value close to one, etc.) it would be useful to get other evidences of nonlinearities and to somehow circumvent the fitting problem. A value of $\beta$ different from one is not only a matter of numerical value, but essentially points to nonlinear effects that are in general relevant at a large scale. In particular, nonlinearities could probably be seen in the dynamics of these systems and this search could constitute an interesting direction for future (urban) studies. Here, we will focus on the much more pragmatic question that lies at the core of the idea of (urban) scaling: knowing some quantity $Y_1$ for a city of size $P_1$ what can we say about the corresponding quantity $Y_2$ for a city of size $P_2$ ? In other words, if we accept the idea of scaling, what is the exponent $\beta$ that we should use in order to compute $Y_2$ according to $Y_2=Y_1(P_2/P_1)^\beta$ ? This simple question is at the core of the analysis presented here. In the next section we present in more details the tools and the method developed here and in the following sections we apply them to different quantities from various datasets.

\section{Scaling: simple tools for a thorough examination}

\subsection{Local exponent across sizes}

We will focus on the `practical' aspect of scaling: instead 
of fitting the data with all the problems discussed above, we consider two cities $1$ and $2$ with populations $P_1$ and $P_2$. Assuming the scaling form Eq.~\ref{eq:pl} to be correct (with the standard assumption of a constant prefactor, see  for example \cite{Bettencourt:2019} for a generalization of the scaling form), knowing $P_1$, $P_2$, and $Y_1$,  we obtain $Y_2$ as
\begin{align}
  Y_2=Y_1\left(\frac{P_2}{P_1}\right)^\beta
  \label{eq:pl2}
\end{align}
The scaling assumption and the value of $\beta$ thus allow us to
predict what will happen to a scaled-up version of a given
city. Conversely, we could also ask what would be the `local' exponent that allows us to predict correctly $Y_2$. Obviously we have from Eq.~\ref{eq:pl2}
\begin{align}
\beta_{\mathrm{loc}}=
\frac
{\log(Y_2/Y_1)}
  {\log(P_2/P_1)}
  \label{eq:betaloc}
\end{align}
which has the simple geometric interpretation of being the slope of the straight line
joining the points $(P_1,Y_1)$ and $(P_2,Y_2)$ in the loglog representation. If there is no noise
and all data points are aligned, we obtain only one value for $\beta_{\mathrm{loc}}$ for all pairs of cities and which also corresponds to the value $\hat{\beta}$ obtained by the direct fit (in the following we will denote the population ratio by $r=P_2/P_1$ where we consider that $P_2$ is the largest population so that we always have $r\geq 1$). In the general case, studying this value $\beta_{\mathrm{loc}}$ tells us how different cities are related to each other giving a representation of scaling across different values of the size ratio, akin to some sort of `tomography' scan of scaling. Plotting $\beta_{\mathrm{loc}}$ versus $r$ is what we will call in this paper the `tomography plot' as it allows us to explore scaling for various cross-sections of the size ratio.

If we assume that $Y_2=Y_1(P_2/P_1)^\beta(1+\eta)$ where $\eta$ is due to noise, we obtain for $P_2/P_1> 1$ the general expression
\begin{align}
\beta_{\mathrm{loc}}=\beta+\frac{\log(1+\eta)}{\log (P_2/P_1)}
\end{align}
This expression shows that when the noise if not too large, the effective exponent converges for large $P_2/P_1$ to the theoretical one and to its estimate via fitting: $\beta_{\mathrm{loc}}\simeq \hat{\beta}\simeq \beta$. This expression also shows that a plot of $\beta_{\mathrm{loc}}$ versus $\log P_2/P_1$ for all pairs of cities should display a hyperbolic envelope and that $\beta_{\mathrm{loc}}\to \beta$ for large size ratio values. For similar populations $P_2=P_1(1+\varepsilon)$ (with $\varepsilon\ll 1$), we obtain at lowest order in $\varepsilon$
\begin{align}
\beta_{\mathrm{loc}}\simeq \beta+\frac{\log(1+\eta)}{\varepsilon}
\end{align} 
We see here that for small $\varepsilon$ we can observe arbitrary large values of $\beta_{\mathrm{loc}}$ for non-zero fluctuations $\eta$. For similar cities, noise is therefore relevant and their comparison cannot help us much in determining the scaling exponent.

In the case of a non-multiplicative noise, we could imagine an expression of the form $Y_2
= Y_1(P_2/P_1)^\beta + \eta$. The noise $\eta$ cannot be too large, otherwise the scaling assumption is 
not correct and $Y_2/Y_1$ won't depend on the ratio $P_2/P_1$ only.   If we however accept
this form, a simple calculation shows that
\begin{align}
  \beta_{\mathrm{loc}}=\beta+\frac{1}{\log
  r}\log(1+\frac{\eta}{r^\beta Y_1})
\end{align}
We thus see that for large $r$ there is a convergence towards $\beta$
for a large class of noise $\eta$. In particular, for $r$ large enough
we have
\begin{align}
  \beta_{\mathrm{loc}}\simeq \beta+\frac{1}{r^\beta\log
  r}\frac{\eta}{Y_1}
\end{align}
which shows that even in this case, if the scaling assumption is
correct, there should be a convergence of $\beta_{\mathrm{loc}}$
towards $\beta$.

We end this part by noting some similarities with multiscaling. Indeed we can rewrite the relation Eq.~\ref{eq:betaloc} as
\begin{align}
  Y_2=Y_1r^{\beta_{\mathrm{loc}}(r)}
\end{align}
and multiscaling is here encoded in the function $\beta_{\mathrm{loc}}(r)$. It is however unclear at this stage if we can connect (and how) the behavior of this function to the existence of multiple scales as it is the case in growth kinetics \cite{Coniglio:1989} and this could constitute an interesting question for future research.

\subsection{Identifying a benchmark city and defining an effective exponent}


This local exponent allows us to define and identify a `benchmark city' that can serve as a reference value for computing quantities for other cities. More precisely, for a city $i$ we first compute the corresponding local exponents for all other cities $j$ versus the ratio $r_{ij}=P_j/P_i$ as \begin{align}
  \beta_{\mathrm{loc}}(i,j)=\frac{\log(Y_j/Y_i)}{\log r_{ij}}
\end{align}
 We then compute the average and the variance of the local exponent when varying $j$
\begin{align}
\langle\beta_{\mathrm{loc}}(i)\rangle&=\frac{1}{N-1}\sum_j\beta_{\mathrm{loc}}(i,j)\\
\sigma^2(i)&=\langle\beta_{\mathrm{loc}}^2(i)\rangle-\langle\beta_{\mathrm{loc}}(i)\rangle^2
\label{eq:variance}
\end{align}
where the brackets denote here $\langle O\rangle=\sum_jO(j)/(N-1)$ ($N$ is the number of cities). We then define the benchmark city such that the variance $\sigma^2(i)$ is the smallest
possible and we denote it by $i_{\min}$. For this city, the fluctuations of the local exponent are the smallest possible around its average $\beta_{\mathrm{eff}}\equiv\langle\beta_{\mathrm{loc}}(i_{\min})\rangle$. This city can then serves as a benchmark in the sense that we can then use it for computing `reliably' properties of other cities through the formula
\begin{align}
Y(j)=Y(i_{\min})\left(\frac{P_j}{P_{i_{\min}}}\right)^{\beta_{\mathrm{eff}}}
\end{align}
and justifies the denomination `effective exponent' as it can be used for practical predictions. Other choices for an effective exponent are of course possible but in the spirit of practical applications we are interested in picking a single value of $\beta$ for computing the quantity $Y$ for all cities. In this respect, minimizing the variance of $\beta_{\mathrm{loc}}$ is a simple sensible answer to this question, although probably not the only one. 

We note here that this discussion is different from the one about SAMIs (Scale-Adjusted Metropolitan Indicators) defined in \cite{Bettencourt:2010,Lobo:2013} as being the variation of a given city with respect to the fit given by $\hat{\beta}$
\begin{align}
  \xi_i=\log\frac{Y_i}{Y_0P_i^{\hat{\beta}}}
\end{align}
We will however consider a similar quantity. Knowing $\beta_{\mathrm{eff}}$, we compute the fraction $f(\varepsilon_1,\varepsilon_2)$ of
cities for which
\begin{align}
\varepsilon_1 Y_{\mathrm{data}}<Y_{\mathrm{predicted}}<\varepsilon_2 Y_{\mathrm{data}}
\end{align}
where $Y_{\mathrm{data}}$ is the actual value for a given city of population $P$ and
\begin{align}
  Y_{\mathrm{predicted}}=Y(i_{min})\left(\frac{P}{P_{i_{min}}}\right)^{\beta_{\mathrm{eff}}}
\end{align}
In particular, we will focus on the case $\varepsilon_1=1/\varepsilon_2$ for different values of $\varepsilon_2$. We will systematically give the value of $f(1/2)$ for $\varepsilon_2=2$ as it gives a good idea of the accuracy of the prediction computed with $\beta_{\mathrm{eff}}$. Additional information can be provided by plotting the function $f(\varepsilon)\equiv f(1/\varepsilon,\varepsilon)$ for $\varepsilon>1$ and we will show it in a few cases.

Finally, we note that it might be possible to construct a more general framework that includes these different definitions and objects, exhibiting possible relations between these tools and we leave this question for future research.

\section{Applications to real-world datasets}

We now apply these tools to the different datasets discussed in \cite{Leitao:2016}. These datasets concern different areas of the world (Europe, USA, OECD, Brazil) and various socio-economical quantities (see table 1) and were analyzed with standard statistical methods. They represent therefore an interesting benchmark dataset for testing other methods. We note that for most of these datasets cities have to be understood as urban areas, except for brazilian datasets where administrative boundaries were used (all the details can be found in \cite{Leitao:2016}). We first discuss clear cases for which there is no or little ambiguity about the scaling behavior and see how it is confirmed with the tools proposed here. We then focus on less clear cases for which we find results not completely consistent with the classical analysis and also on the datasets where the statistical analysis in \cite{Leitao:2016} was `inconclusive', meaning that the result was depending on the assumption taken for the disorder. Our main goal here will be to show how our tools can shed a new light on these problematic or inconclusive cases.

\subsection{Simple cases}

\subsubsection{Income and patents in the UK}

\begin{figure}[ht!]
\includegraphics[width=0.9\linewidth]{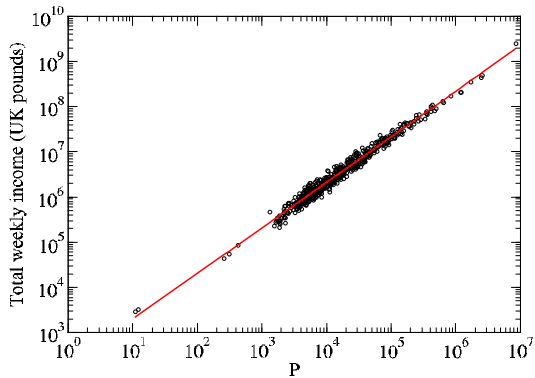}
\caption{ Total weekly income for cities in the UK (see \cite{Leitao:2016} and \cite{Arcaute:2015} for a description of the data) versus population. The power law fit gives $\beta=1.01$ and $r^2=0.99$.}
\label{fig:income}
\end{figure}

We first consider the case of the total weekly income in the UK considered in \cite{Leitao:2016}. In this case the behavior seems to be
linear and the fit gives $\hat{\beta}=1.01$ ($r^2=0.99$). We note here that this result is consistent with those found in \cite{Arcaute:2015} for various definitions of cities. The fit is shown in Fig.~\ref{fig:income}.

\begin{figure}[ht!]
\includegraphics[width=0.9\linewidth]{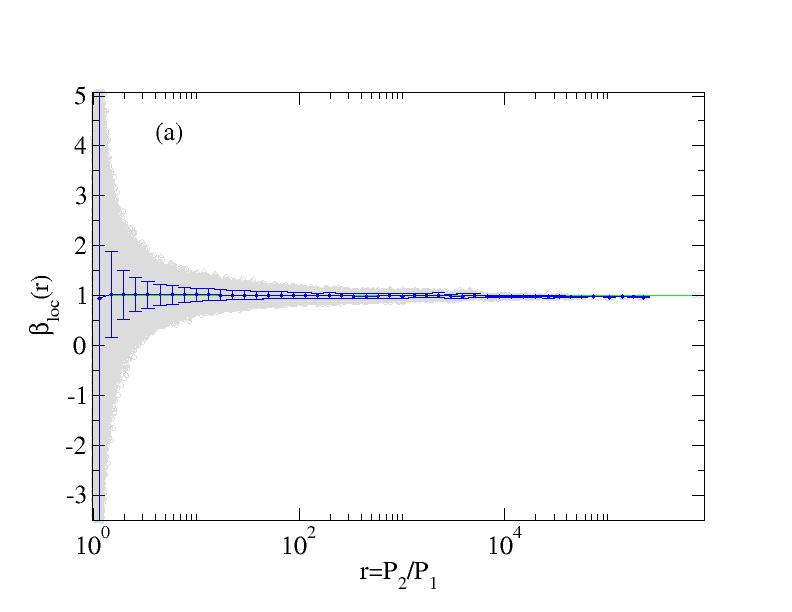}
\includegraphics[width=0.9\linewidth]{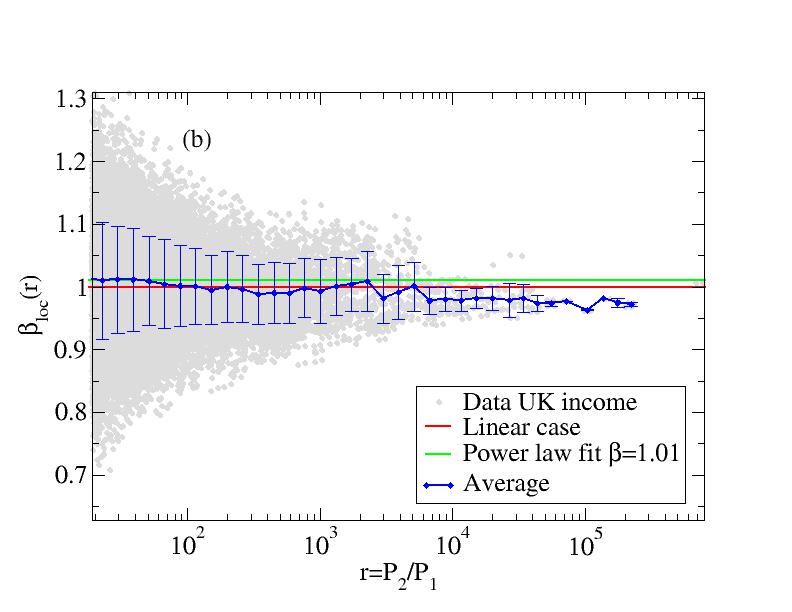}
\caption{(a) `Tomography plot' for the total weekly income in UK cities: local exponent $\beta_{loc}$ for the income in the UK versus the size ratio $r=P_2/P_1$. (b) Same as in (a) but zoomed on the large $P_2/P_1$ region. The error bars here and in the following correspond to 1 standard deviation computed for each $r$ bin. } 
\label{fig:incomemueff}
\end{figure}

We compute the local exponent $\beta_{\mathrm{loc}}$ versus $P_2/P_1$ in this case and obtain the result shown in Fig.~\ref{fig:incomemueff}. We also show the average of $\beta_{\mathrm{loc}}$ in each $r-$bin and the corresponding error bar computed as the standard dispersion, the values corresponding to the linear case ($\beta=1$) and to the power law fit $\hat{\beta}=1.01$).
We observe on this plot large fluctuations for small  $r=P_2/P_1$ as expected: in this regime the local exponent is governed
by fluctuations among cities of similar population sizes. For larger $r$ we observe a quick convergence to $1$, and for most pairs of cities with $r\gtrsim 100$ we observe $\beta_{\mathrm{loc}}=1.0\pm 0.1$. We also note that for very large ratio  $r\gtrsim 10^4$, the local exponent is slightly smaller than one (Fig.~\ref{fig:incomemueff}b).

\begin{figure}[ht!]
  \includegraphics[width=0.9\linewidth]{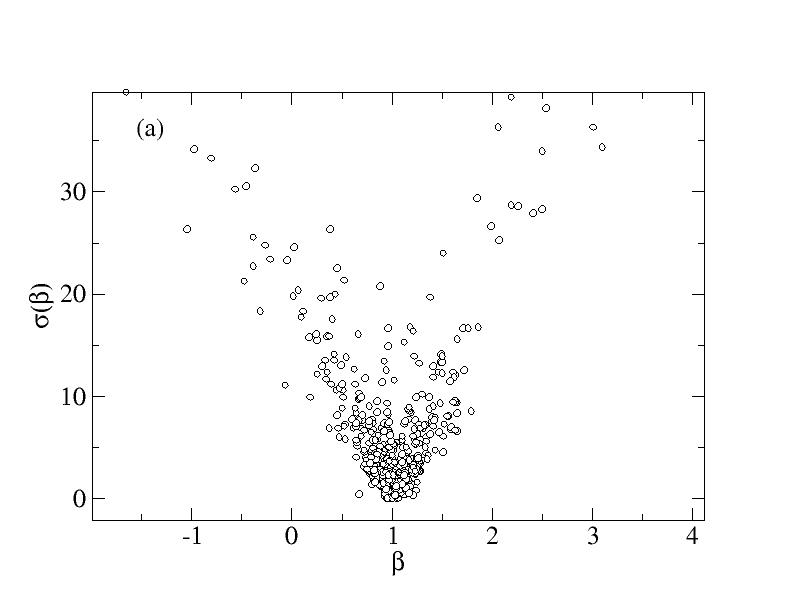}\\
  \includegraphics[width=0.9\linewidth]{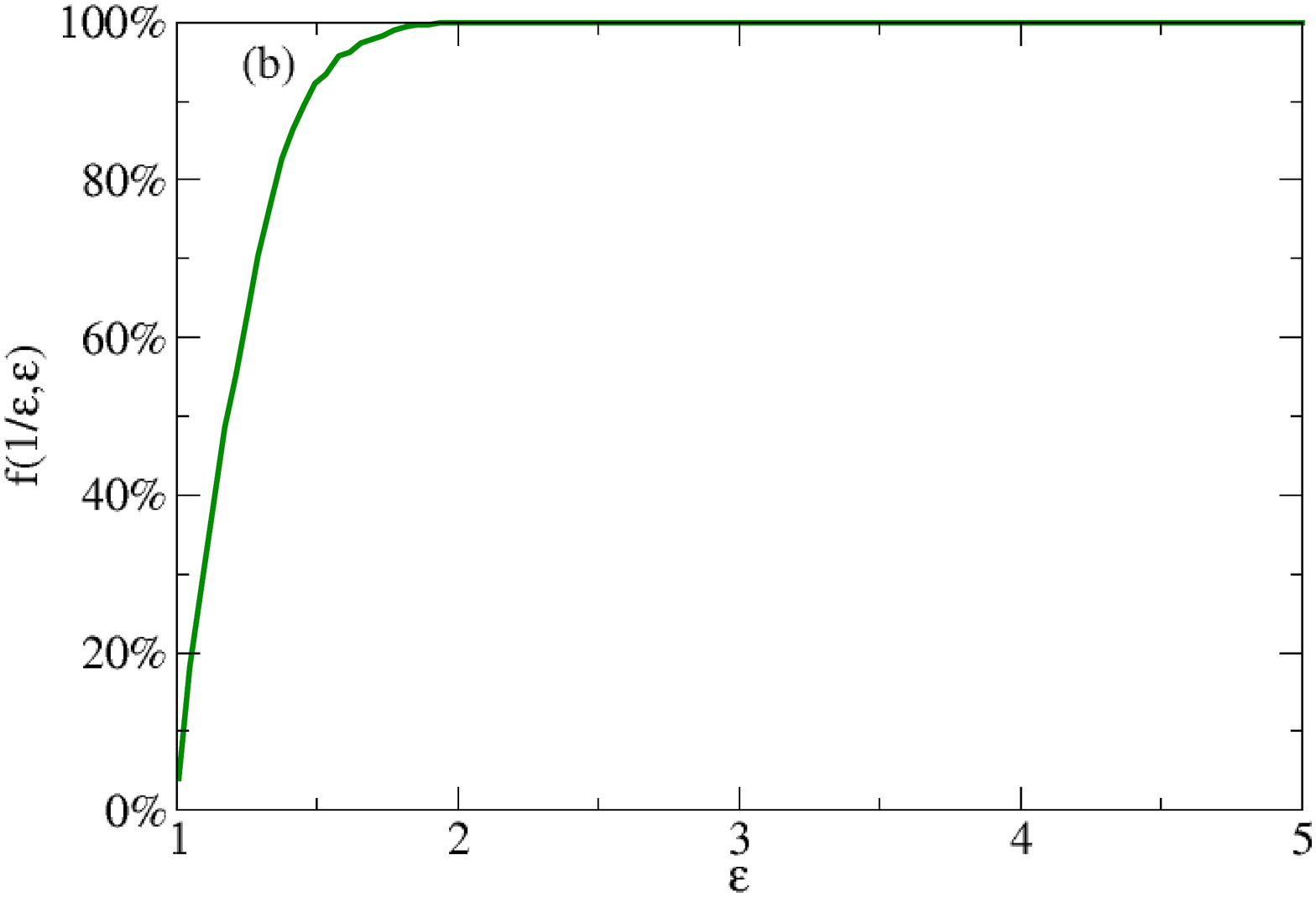}\\
    \includegraphics[width=0.9\linewidth]{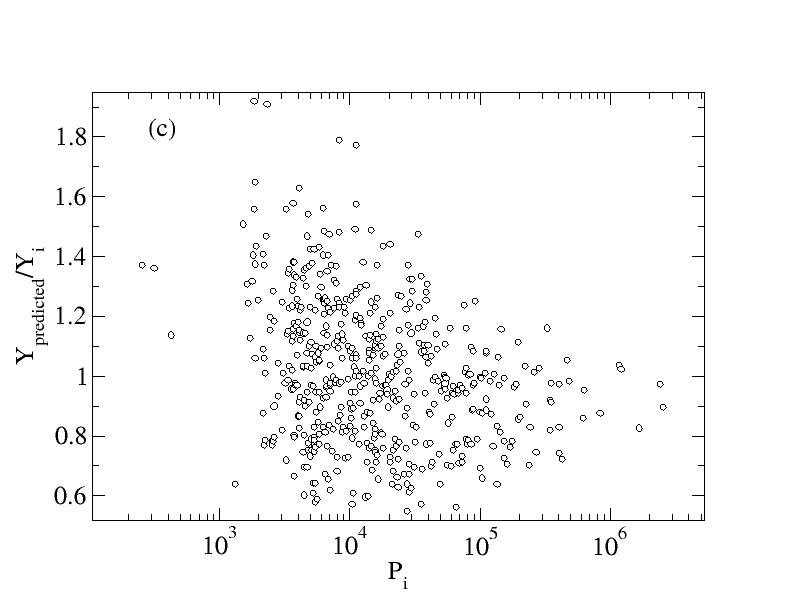}
    \caption{(a) Dispersion $\sigma$ versus the value of the exponent $\beta$ (we show here the zoom on the part with $\sigma<40$). (b) Variation of $f(1/\varepsilon,\varepsilon)$ with $\varepsilon>1$ for $\beta=\beta_{\mathrm{eff}}$. We observe that for $\varepsilon>2$, the ratio $Y_{predicted}/Y_{data}$ is at least in the range $[0.5,2.0]$ for all cities. (c) Using the benchmark city, we compare predictions and data with the ratio $Y_{predicted}/Y_{data}$. We observe here that for all pairs of cities this ratio is in the range $[0.5,2.0]$.}
\label{fig:benchincome}
\end{figure}

In order to complete this picture we now identify the `benchmark' city, defined above as the city which allows the most reliable prediction (i.e. with the smallest fluctuations), and obtained as the minimization of the variance given by Eq.~\ref{eq:variance}. The `effective' exponent is the one associated to this benchmark city as it can be used for reliable predictions. Instead of plotting $\beta$ and $\sigma$ for each city, we directly show on Fig.~\ref{fig:benchincome}a the dispersion $\sigma$ versus $\beta$ and we find that it is the smallest city of the dataset that is the benchmark. The corresponding effective exponent is $\beta_{\mathrm{eff}}=0.97\pm 0.03$.
For this value of $\beta_{\mathrm{eff}}$, we study the evolution of the function $f(\varepsilon)$ which represents the fraction of cities for which the prediction lies in the range $[Y_{data}/\varepsilon, Y_{data}\varepsilon]$ and show it on the Fig.~\ref{fig:benchincome}(b).  We also display on the Fig.~\ref{fig:benchincome}(c) the ratio $Y_{predicted}/Y_{data}$ versus the population $P_i$ and which shows that the fraction $f(1/2)$ of cities with a ratio $Y_{predicted}/Y_{data}$ in the range $[0.5,2.0]$ is $f(1/2)=100\%$ (we note here that for another value such as $\beta=\hat{\beta}$ the corresponding value is a bit smaller $f(1/2)=97\%$).

We thus see on this example a convergence of evidences: the naive fit gives $\hat{\beta}\approx 1$, the quantity $\beta_{\mathrm{loc}}$ converges quickly towards $1$ and most pairs of cities are related to each other via a linear relation. Finally, the most reliable way to compute the income for a city $i$ is to use an effective exponent $0.97$ demonstrating a slight sublinearity (which comes from pairs of nodes with a large population ratio as can be seen in Fig.~\ref{fig:incomemueff}b).

\begin{figure}[ht!]
  \includegraphics[width=0.9\linewidth]{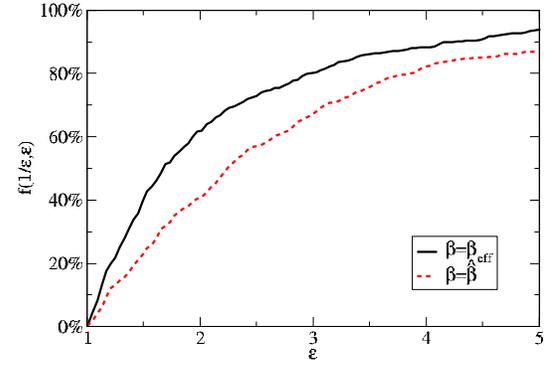}
  \caption{Patents in the UK: Variation of $f(1/\varepsilon,\varepsilon)$ with $\varepsilon>1$ for $\beta=\beta_{\mathrm{eff}}$ and $\beta=\hat{\beta}$. We observe here that in terms of prediction the value $\beta_{\mathrm{eff}}$ is a better choice than the value obtained by fitting.}
\label{fig:patentsfepsilon}
\end{figure}

We observe the same sort of behavior for patents in the UK (plots not shown): the fit gives $\hat{\beta}=1.06$ ($r^2=0.88$), an effective exponent $\beta_{\mathrm{eff}}=0.96$. We compare for this case the functions $f(1/\varepsilon,\varepsilon)$ obtained in the different cases $\beta=\beta_{\mathrm{eff}}$ and $\beta=\hat{\beta}$ (see Fig.~\ref{fig:patentsfepsilon}).
We see that the effective exponent is always better in terms of predictions, with in particular a value $f(1/2)=62\%$ consistent with a linear behavior. We note that this linear result for the patents in the UK is in contrast with the result obtained for the US in \cite{Bettencourt:2007a} and more recently in \cite{Bettencourt:2010} with a nonlinear behavior characterized by $\beta=1.28$ while in \cite{Arcaute:2015}  the results for patents in UK cities seem to strongly depend on the definition of cities. We will discuss in more detail this case of US patents below in the `Problematic cases' section.

\subsubsection{Clear nonlinear behavior: the USA case}

\begin{figure}[ht!]
  \includegraphics[width=0.9\linewidth]{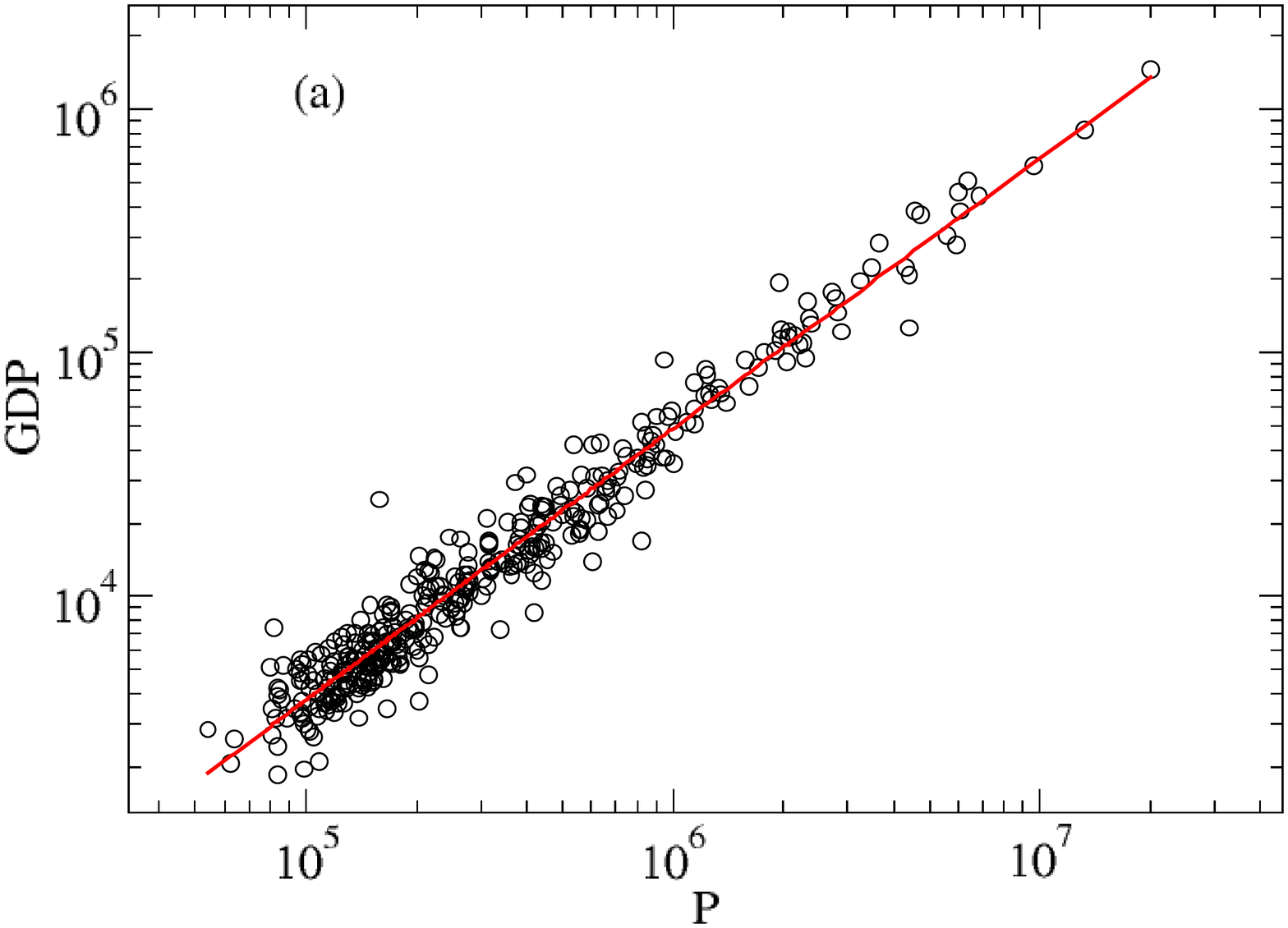}
  \includegraphics[width=0.9\linewidth]{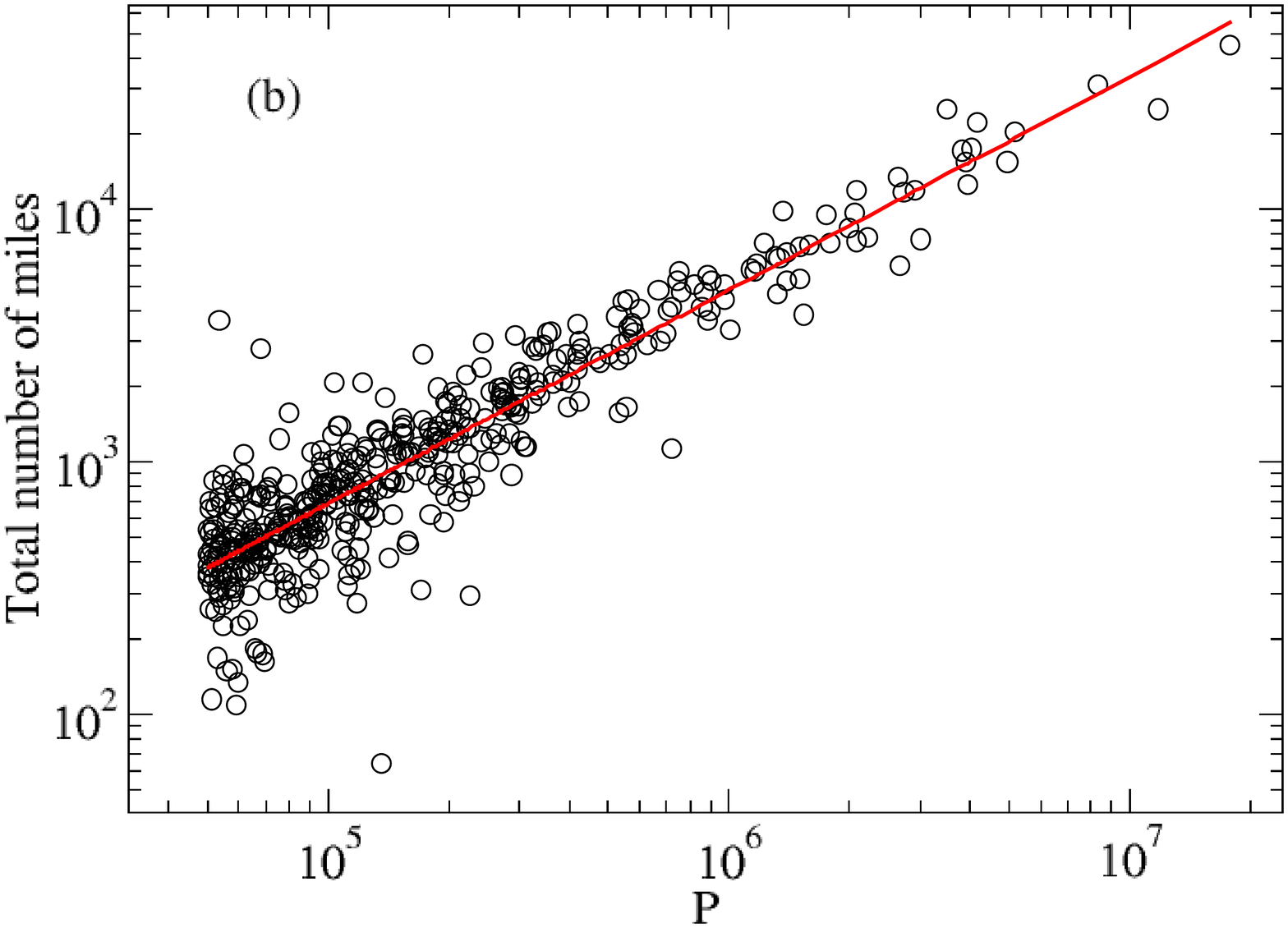}
  \caption{(a) GDP (in millions of current US dollars) for cities in the USA for the year 2013 \cite{Leitao:2016}. The red line is a power law fit with $\hat{\beta}=1.11$ and $r^2=0.98$. (b) Number of miles for US cities (year 2013) versus population. The power law fit gives $\beta=0.85$ ($r^2=0.91$).}
\label{fig:usafit}
\end{figure}

\begin{figure}[ht!]
  \includegraphics[width=0.9\linewidth]{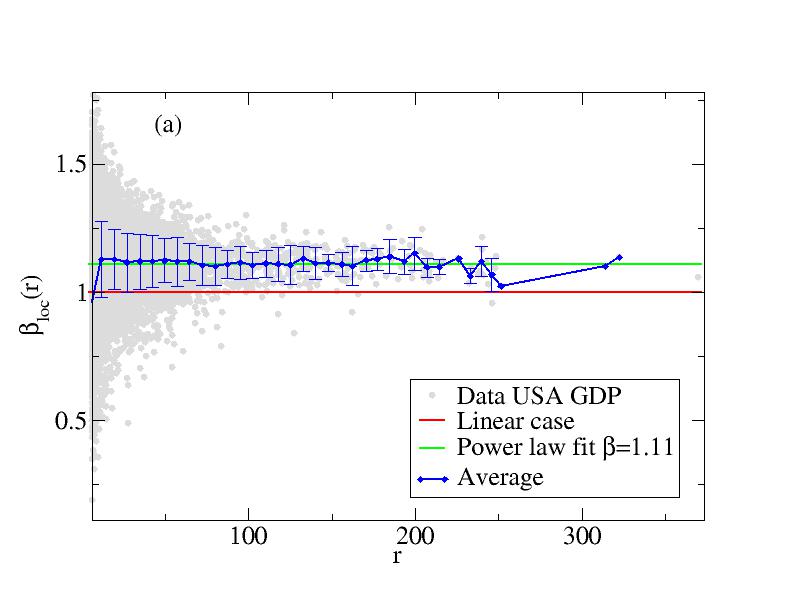}
  \includegraphics[width=0.9\linewidth]{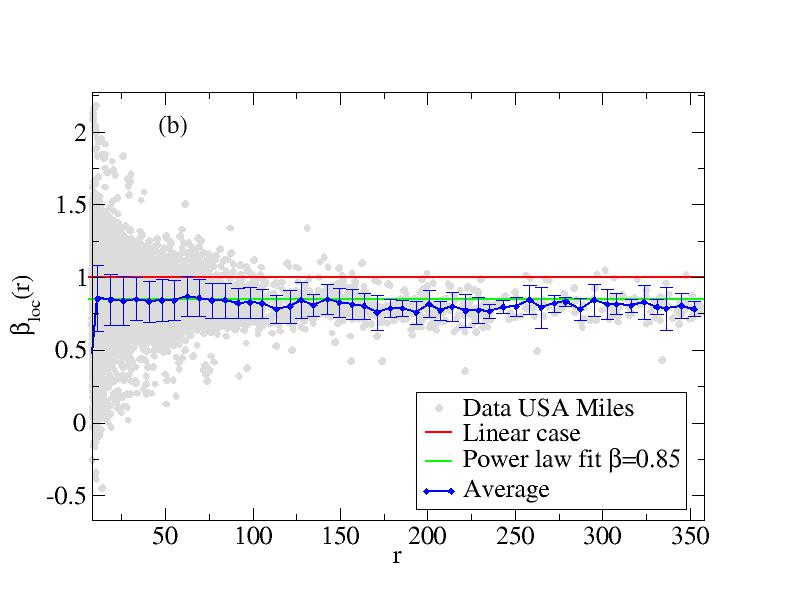}
  \caption{Local exponent versus population ratio $r$ for (a) the GDP for cities in the USA (year 2013) and (b) the total number of miles in the USA (year 2013).}
\label{fig:usatomo}
\end{figure}

We now consider here the two datasets for the USA that were studied in \cite{Leitao:2016}. The first one is about the GDP of cities and the second one about the number of miles of roads (in each city). The nonlinear fits for these two quantities are shown in Fig.~\ref{fig:usafit}.
In the first case the GDP displays a clear superlinear behavior with $\hat{\beta}=1.11$, while for infrastructure the expected sublinear behavior is observed with $\hat{\beta}=0.85$. We now inspect in more detail these cases with the help of the local exponent $\beta_{\mathrm{loc}}$ (see Fig.~\ref{fig:usatomo}).
We observe on these plots that the `naive' nonlinear fitting is confirmed: for most pairs of cities, the local exponent is different from one and is equal to $\hat{\beta}$ (within error bars). If we now compute the effective exponents we obtain for the GDP $\beta_{\mathrm{eff}}=1.13\pm 0.07$ and for the number of miles $\beta_{\mathrm{eff}}=0.80\pm 0.1$. We note that in both cases the benchmark city is New York city, the largest urban area in the US (at this point we note that there seems to clear rule for identifying the benchmark city). We see here that all evidences are pointing to the same conclusion of a nonlinear behavior. Even if $\beta=1.11$ is only slightly different from $1$ the tomography plot (Fig.~\ref{fig:usatomo}a) clearly shows the reality of this nonlinear exponent and is confirmed by the value of the effective exponent $\beta_{\mathrm{eff}}=1.13$. Using these effective exponents for the GDP and the number of miles, the fraction $f(1/2)$ is $98\%$ and $91\%$ for the GDP and the miles, respectively. In other words, using NYC as the benchmark city and the effective exponents, we get excellent predictions for all the other cities. 

\begin{figure}[ht!]
    \includegraphics[width=0.9\linewidth]{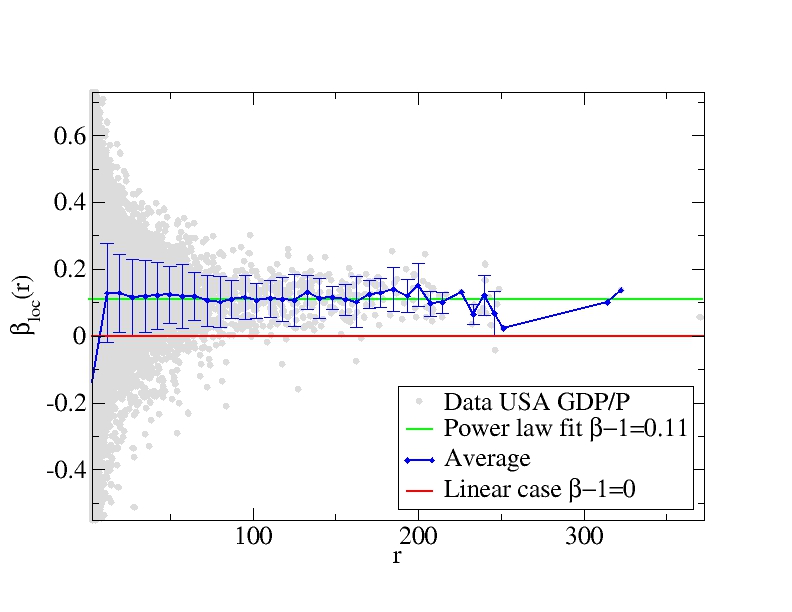}
  \caption{Tomography plot for the GDP per capita for US cities (year 2013).}
\label{fig:usatomo2}
\end{figure}

Finally, in order to address the problem discussed in \cite{Shalizi:2011}, we redo the analysis in the US case for the GDP per capita. In this case the nonlinear fit is indeed less good with an exponent $0.11\;(r^2=0.42)$ but which corresponds well to the value $\hat{\beta}-1$. The tomography plot constructed for this quantity is shown in Fig.~\ref{fig:usatomo2} and seems to be free of any ambiguity: for most values of $r$ the local exponent is equal on average to $0.11$ and converges quickly towards this value when $r$ increases, and is strictly positive (within error bars). In addition the effective exponent is in this case equal to $\beta_{\mathrm{eff}}=0.13\pm 0.07$ and leads to an impressive value of the fraction of cities $f(1/2)=98\%$ whose value is correctly predicted. All these elements suggest that there is indeed a nonlinear behavior for the GDP in US cities, even if we work on the GDP per capita that should exclude effects due to the extensivity of this quantity as claimed in \cite{Shalizi:2011}.

\subsubsection{Nonlinear behavior with large fluctuations: museum and libraries in Europe}

\begin{figure}[ht!]
  \includegraphics[width=0.9\linewidth]{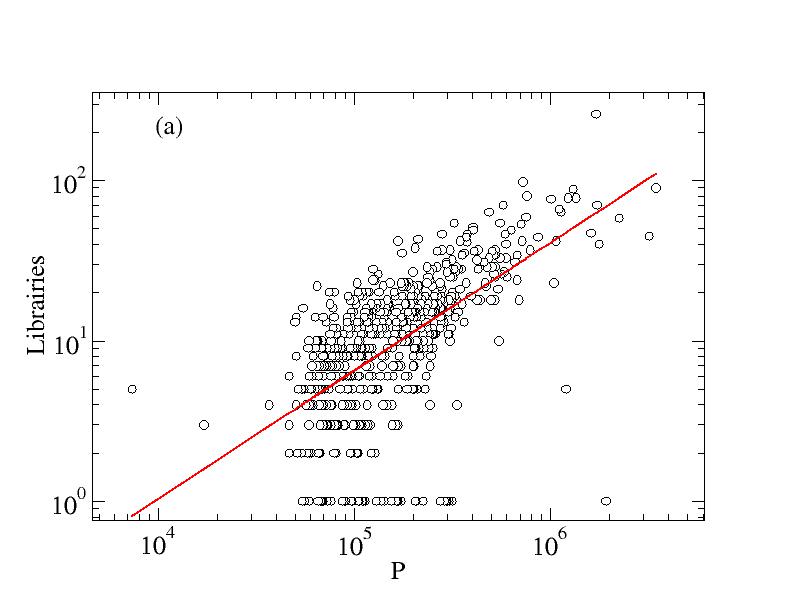}
      \includegraphics[width=0.9\linewidth]{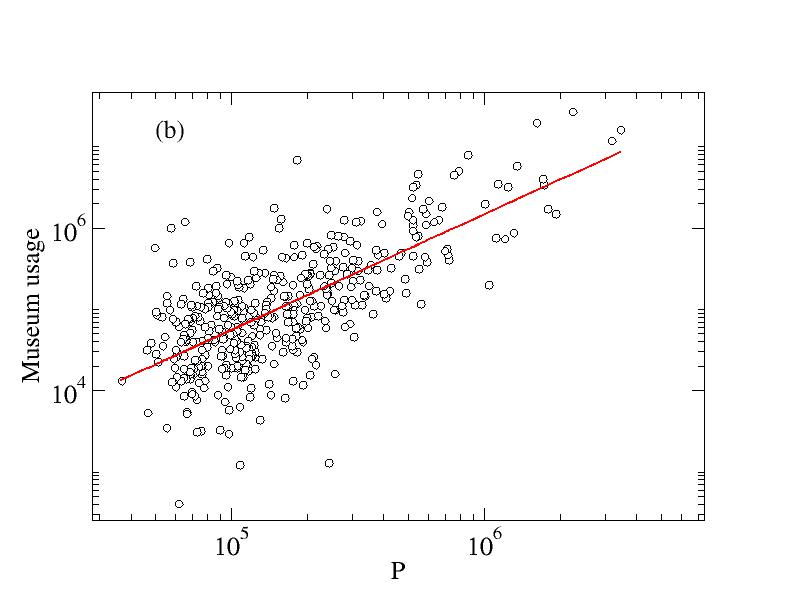}
      \caption{(a) Number of libraries in European cities (2011). The red line is a power law fit with $\hat{\beta}=0.80\;(r^2=0.59)$. (b) Yearly attendance of museums in European cities (2011). The red line is a power law fit with $\hat{\beta}=1.42\;(r^2=0.69)$.}
\label{fig:eurocult2}
\end{figure}

\begin{figure}[ht!]
  \includegraphics[width=0.9\linewidth]{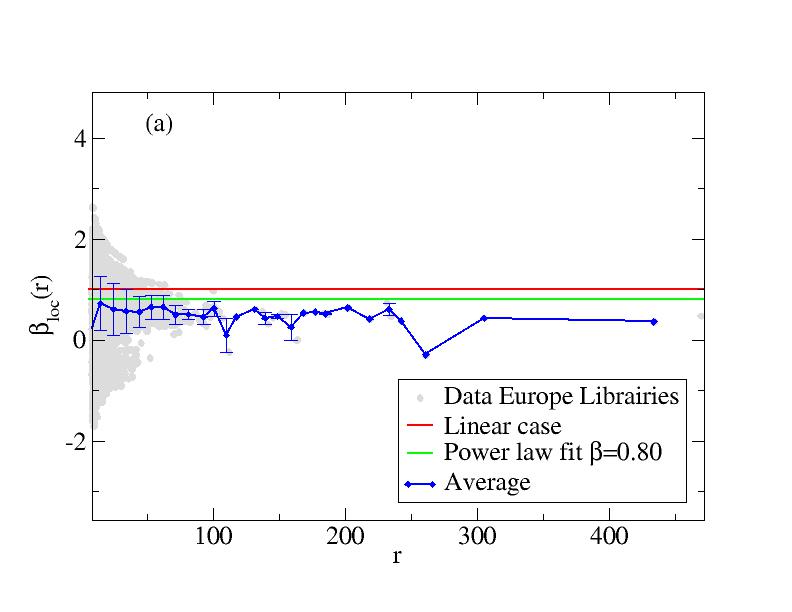}
  \includegraphics[width=0.9\linewidth]{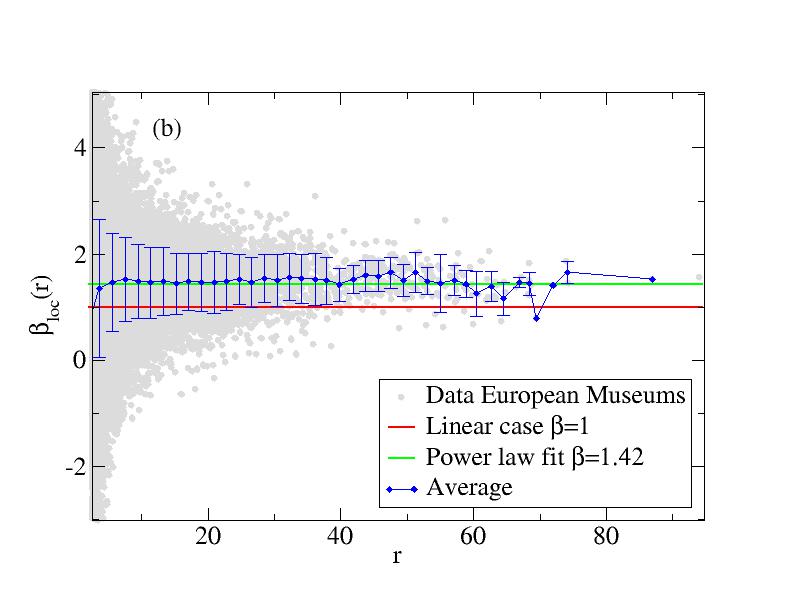}
  \includegraphics[width=0.9\linewidth]{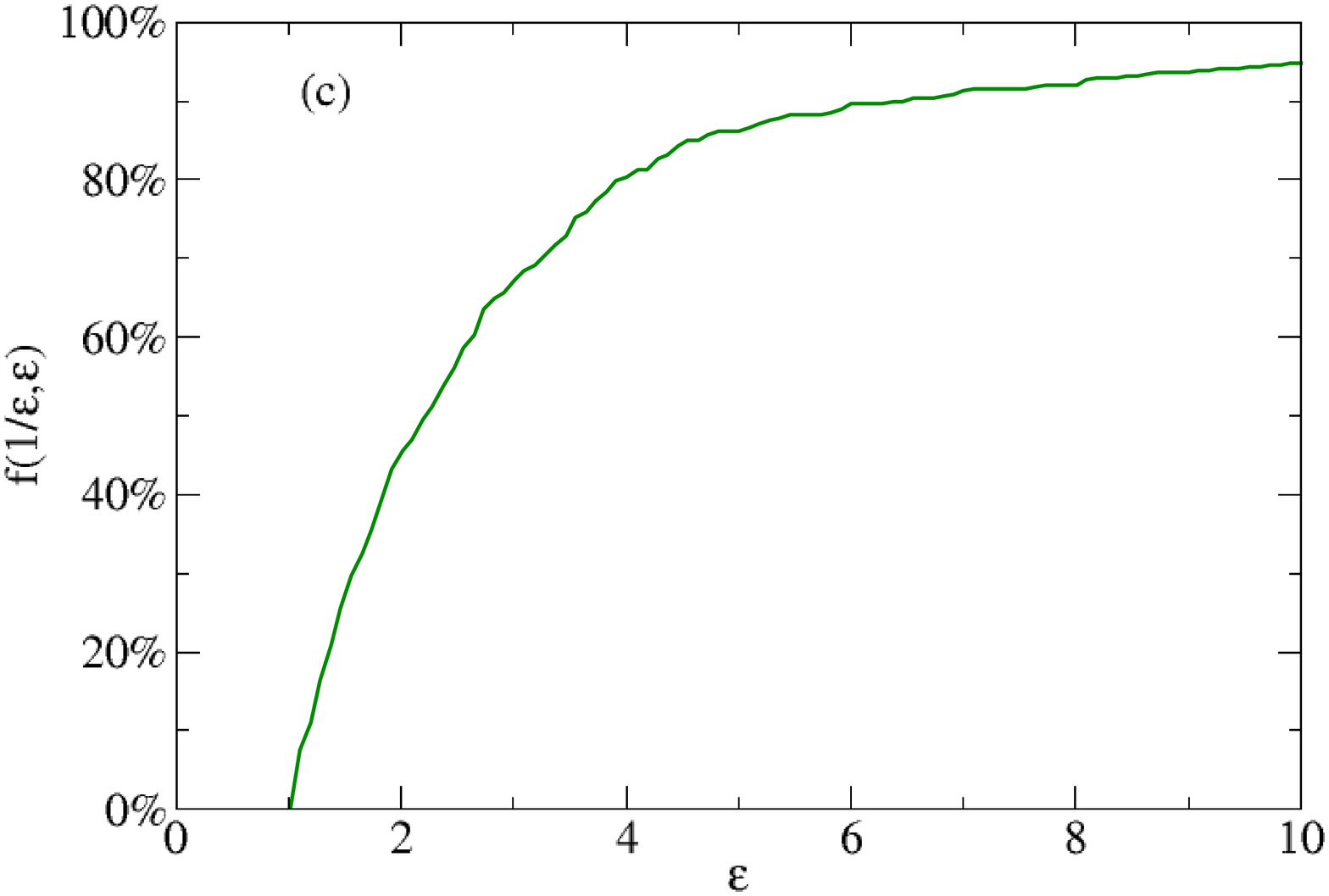}
  \caption{Tomography plots for (a) the number of libraries in European cities (2011), and (b) the yearly attendance of museums in European cities (2011). (c) Fraction $f(1/\varepsilon,\varepsilon)$ versus $\varepsilon$ for the museum usage in European cities.}
\label{fig:eurocultomo}
\end{figure}

\begin{figure}[ht!]
    \includegraphics[width=0.9\linewidth]{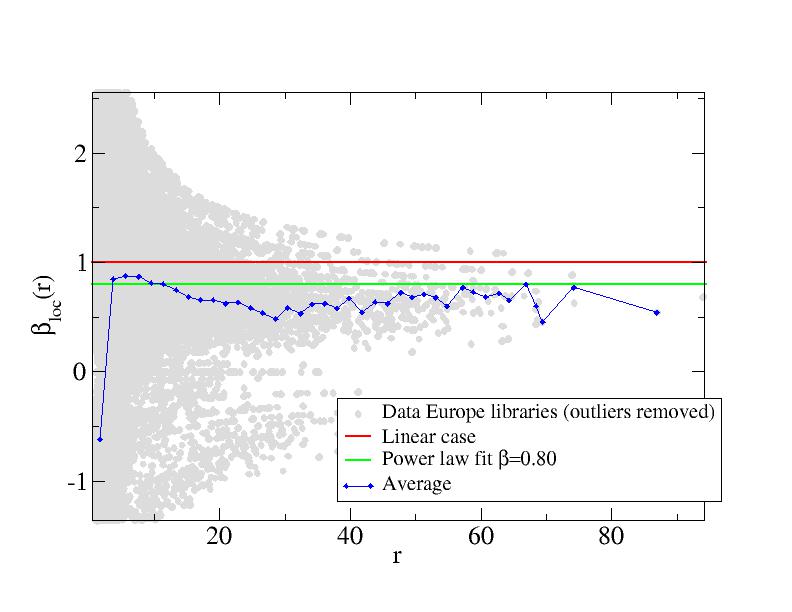}
  \caption{Tomography plot for the number of libraries in European
    cities with outliers
    removed ($P<10^4$ or $Y>10^2$).}
\label{fig:rmout}
\end{figure}

We now consider two other datasets that were also studied in \cite{Leitao:2016}: the attendance of museums (in the year 2011), and the number of public libraries in each city (also for the year 2011). The authors found that there is a superlinear behavior for the museum case, and a sublinear one for the number of libraries. Both these results are sensible: we expect that the number of libraries scales sublinearly with population as it would be the case for many other facilities \cite{Barthelemy:2016}. Also, it is not difficult to accept that the attendance of museums can largely benefit from positive interaction effects in cities, leading to a superlinear behavior. These quantities versus populations are shown in the Fig.~\ref{fig:eurocult2}.
In both cases, we observe that there are large fluctuations and not much more than one decade over which the fit is made. For libraries the power law fit gives $\hat{\beta}=0.80\;(r^2=0.59)$ and for museum usage $\hat{\beta}=1.42\;(r^2=0.69)$. 
The tomography plots for these cases are shown in Fig.~\ref{fig:eurocultomo},
and provide further information. First for libraries, there is no convergence of $\beta_{\mathrm{loc}}$ towards $\hat{\beta}$ consistent with the fact that the power law fit is indeed not reliable. In addition, we find that $\beta_{\mathrm{eff}}=0.17\pm0.32$ (very different from $\hat{\beta}$) and $f(1/2)=55\%$ showing that even the most reliable power law exponent accounts for about half of the data only. This is a case where our analysis actually weakens the conclusions obtained with standard statistical tools. The situation is obviously improved if we remove outliers. For example, if we remove cities with small population ($P<10^4$) or with a large number of libraries ($Y>10^2$), the naive fit remains the same with
a value $\hat{\beta}=0.807$ ($r^2=0.73$), and the tomography plot is
  shown in Fig.~\ref{fig:rmout}.
We observe that the range of x-axis is obviously smaller (as we
removed small size cities), but that the qualitative behavior remains
the same, namely with a relative consistency towards a sublinear
behavior. As expected, the sublinear behavior is better supported here  as outliers
are removed.

The situation is very different for museum usage as shown in Fig.~\ref{fig:eurocultomo}b: there is a clear convergence of $\beta_{\mathrm{loc}}$ towards $\hat{\beta}=1.42$, a superlinear behavior confirmed by $\beta_{\mathrm{eff}}=1.64\pm 0.47$, but with $f(1/2)=45\%$ signalling the presence of very large fluctuations. We also see the effect of these large fluctuations in the slowly increasing fraction $f(1/\varepsilon,\varepsilon)$ for increasing $\varepsilon>1$ shown in Fig.~\ref{fig:eurocultomo}(c). We thus see on these two examples  how our analysis can bring further insights about the quality of the fit.

\subsection{Problematic cases}

We consider here datasets for which the analysis in \cite{Leitao:2016} 
didn't apparently pose too many problems but for which our tools revealed some difficulties. These datasets are the UK railroads, AIDS cases in Brazil, and the number of patents in cities belonging to OECD countries. 

\subsubsection{UK Railroads and AIDS cases in Brazil: existence of a threshold}

\begin{figure}[ht!]
  \includegraphics[width=0.9\linewidth]{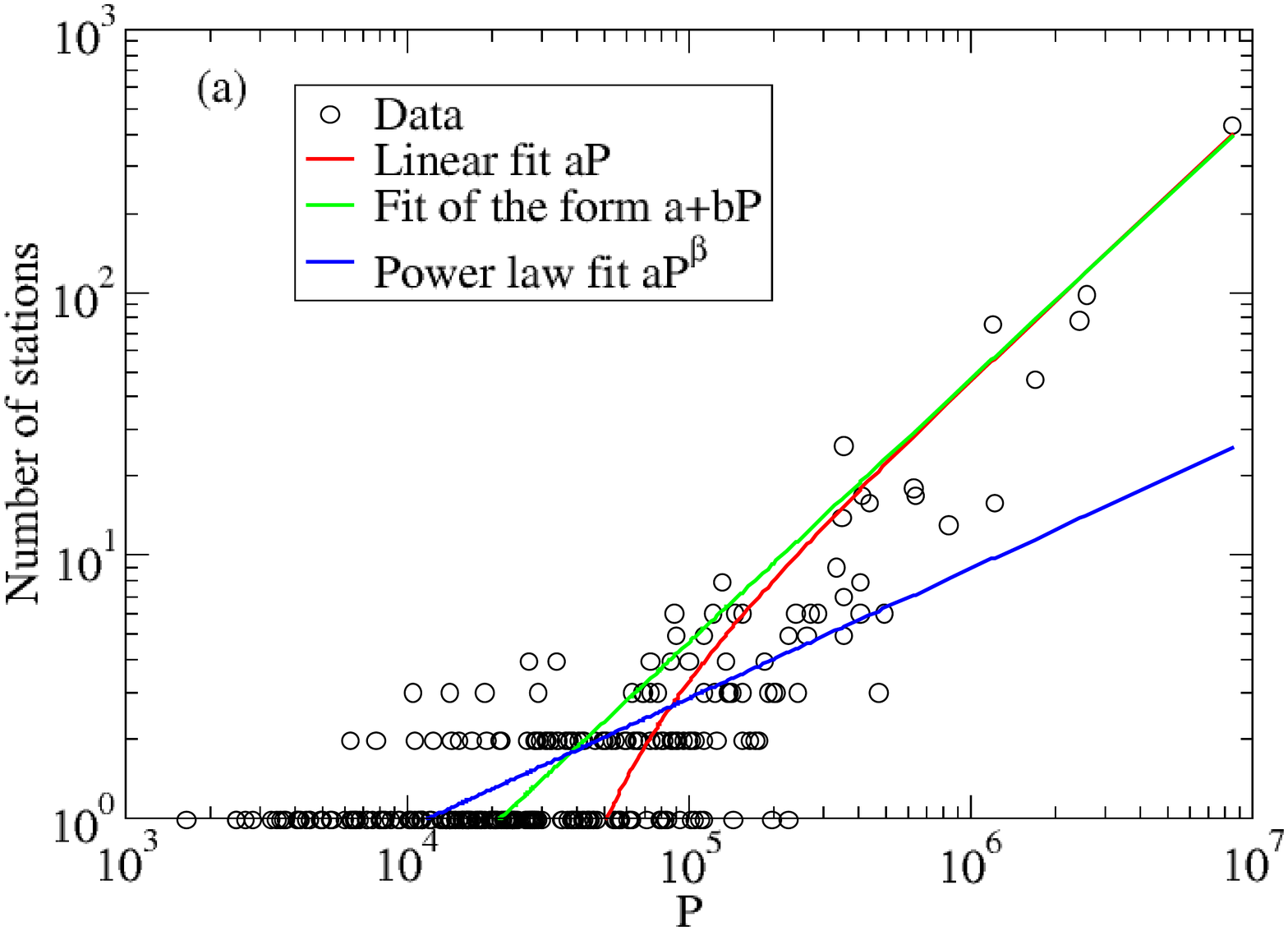}\\
  \includegraphics[width=0.9\linewidth]{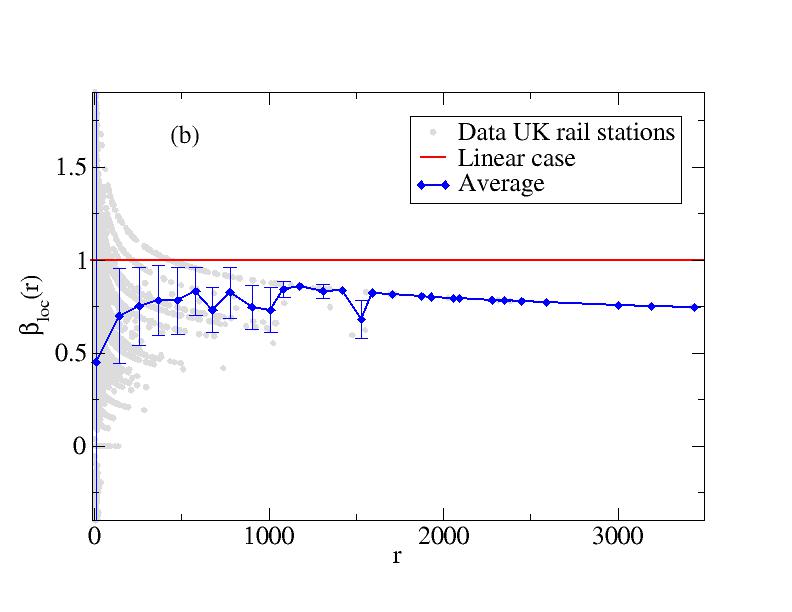}
  \caption{(a) Number of rail stations vs. population in UK cities \cite{Leitao:2016}. We show here the linear fit $aP$ with $a\approx 4.67 10^{-5}$ ($r^2=0.98$), the fit $a+bP$ where $a\approx -1.42$ and $b\approx 4.72 10^{-5}$ ($r^2=0.98$), and the power law fit $aP^\beta$ where $a\approx 0.01$ and $\beta\approx 0.50$ ($r^2=0.76$). (b) Tomography plot for the number of rail stations. }
\label{fig:ukrail}
\end{figure}

The authors of \cite{Leitao:2016}  studied the number of train stations in UK cities and found
a linear behavior $\hat{\beta}=1.0$. However, if we plot this number versus the population 
we obtain the result shown in Fig.~\ref{fig:ukrail}a.
We first observe that there is a lot of noise and the quality of any fit will likely be very poor. Also, we note that there is a large number of cities with one station exactly which potentially will impact any fitting method. Given all these problems, the linear fit is not too bad, in agreement with the result of the analysis of \cite{Leitao:2016}. However the plot of the local exponent versus $r$ shown in Fig.~\ref{fig:ukrail}b signals the existence of important problems. Indeed this plot seems to indicate a sublinear behavior, far from the linear prediction, but also with very large fluctuations (the different hyperbolas appears because of cities with the same number of stations such as 1, or 2 stations. etc. -- we added noise to the data in order to destroy this effect and observe that the tomography plot is robust). This inconsistency suggests the existence of  a problem in this dataset. The presence of large fluctuations could be a reason for the discrepancy observed between the linear behavior and $\beta_{loc}(r)$, but it could also signal another scaling form. In particular, the data is not inconsistent with a fit of the form $a+bP$ where $a<0$ (see Fig.~\ref{fig:ukrail}a) implying a threshold effect: for $P<P_c\approx 30,084$ we have no stations while for $P\gg P_c$ we observe a linear behavior. In the power law scaling assumption, we can compute the effective exponent and find $\beta_{\mathrm{eff}}=0.12\pm 0.17$ with $f(1/2)=85\%$, but given the large level of noise and the high likelihood of another scaling form, we don't assign a high confidence in this result. 

\begin{figure}[ht!]
 \includegraphics[width=0.9\linewidth]{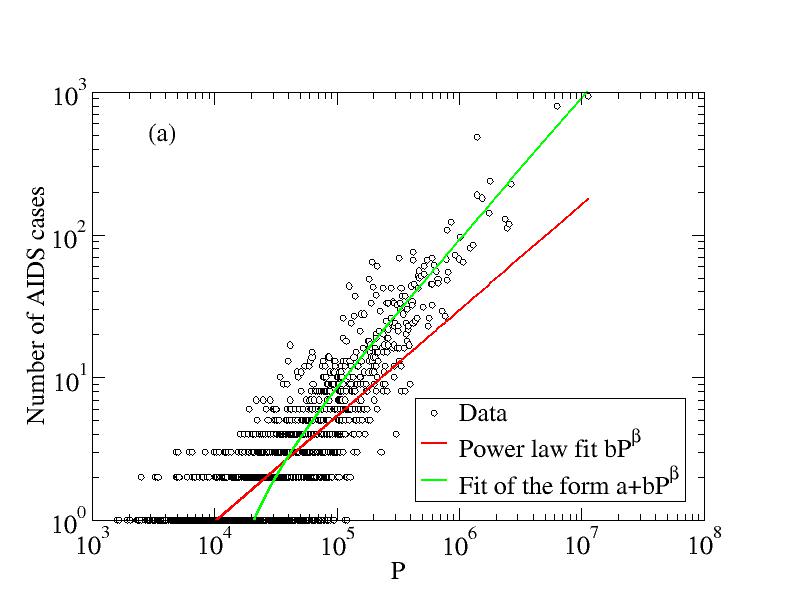}\\
  \includegraphics[width=0.9\linewidth]{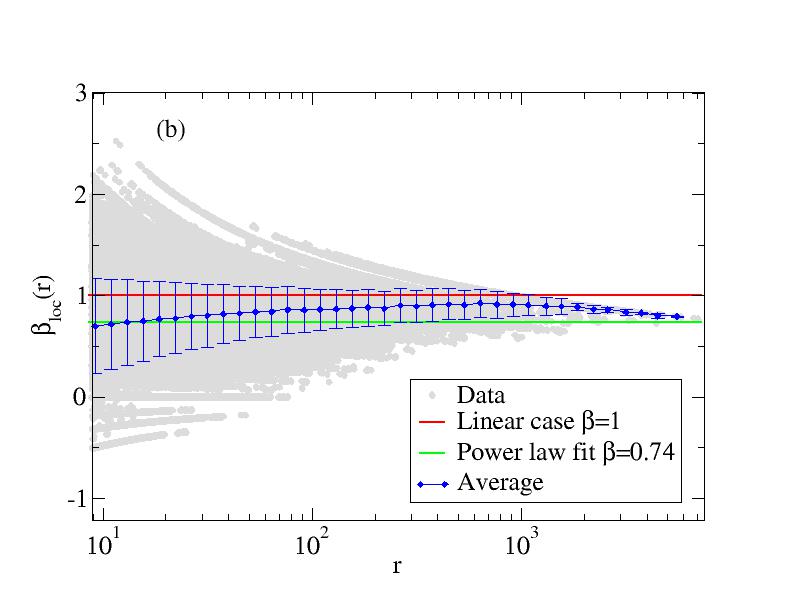}
  \caption{(a) Number of AIDS cases versus city population in Brazil (for 2010). We show both the power law fit 
    s the power law fit with exponent $\hat{\beta}=0.74$ ($r^2=0.81$) and the fit of the form $a+bP^{\hat{\beta}}$ with $\hat{\beta}=0.99$, $a=-1.009$, and $b=0.00010$ ($r^2=0.93$). (b) Corresponding tomography plot.}
\label{fig:brazilaids}
\end{figure}

The situation for the number of AIDS cases in Brazil (for the year 2010) is similar to the previous case. The plot of this number versus population is shown in Fig.~\ref{fig:brazilaids}a.
The power law fit gives an exponent $\hat{\beta}=0.74$ consistent with the sublinear conclusion of \cite{Leitao:2016}, but given the large fluctuations a fit of the form $a+bP$ is also consistent with the data. This last fit predicts a threshold effect with $P_c=10,090$ and a linear behavior for $P\gg P_c$, similarly to the previous case of UK rail stations. The tomography plot (Fig.~\ref{fig:brazilaids}b) shows that the scaling behavior is not clear with a range around $r\sim 10^3$ for which the local exponent is close to $1$ but for other values of $r$ we observe a sublinear exponent. The effective exponent is $\beta_{\mathrm{eff}}=1.03$ with a fraction $f(1/2)=67\%$. It thus seems here that the sublinear conclusion of \cite{Leitao:2016} could actually be challenged by a threshold function and/or  a linear behavior.

\subsubsection{OECD Patents: not a simple scaling function ?}

\begin{figure}[ht!]
  \includegraphics[width=0.9\linewidth]{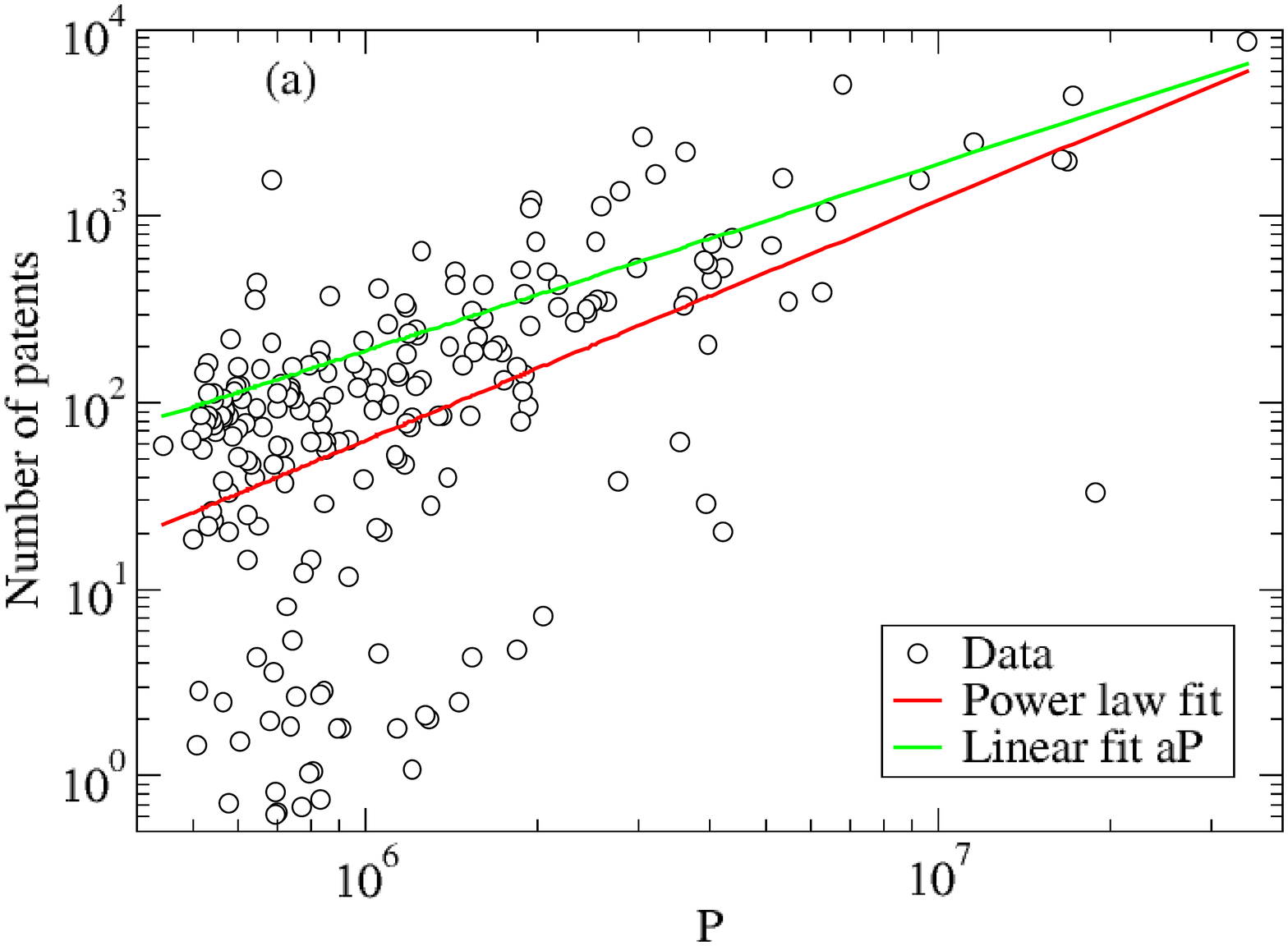}\\
  \includegraphics[width=0.9\linewidth]{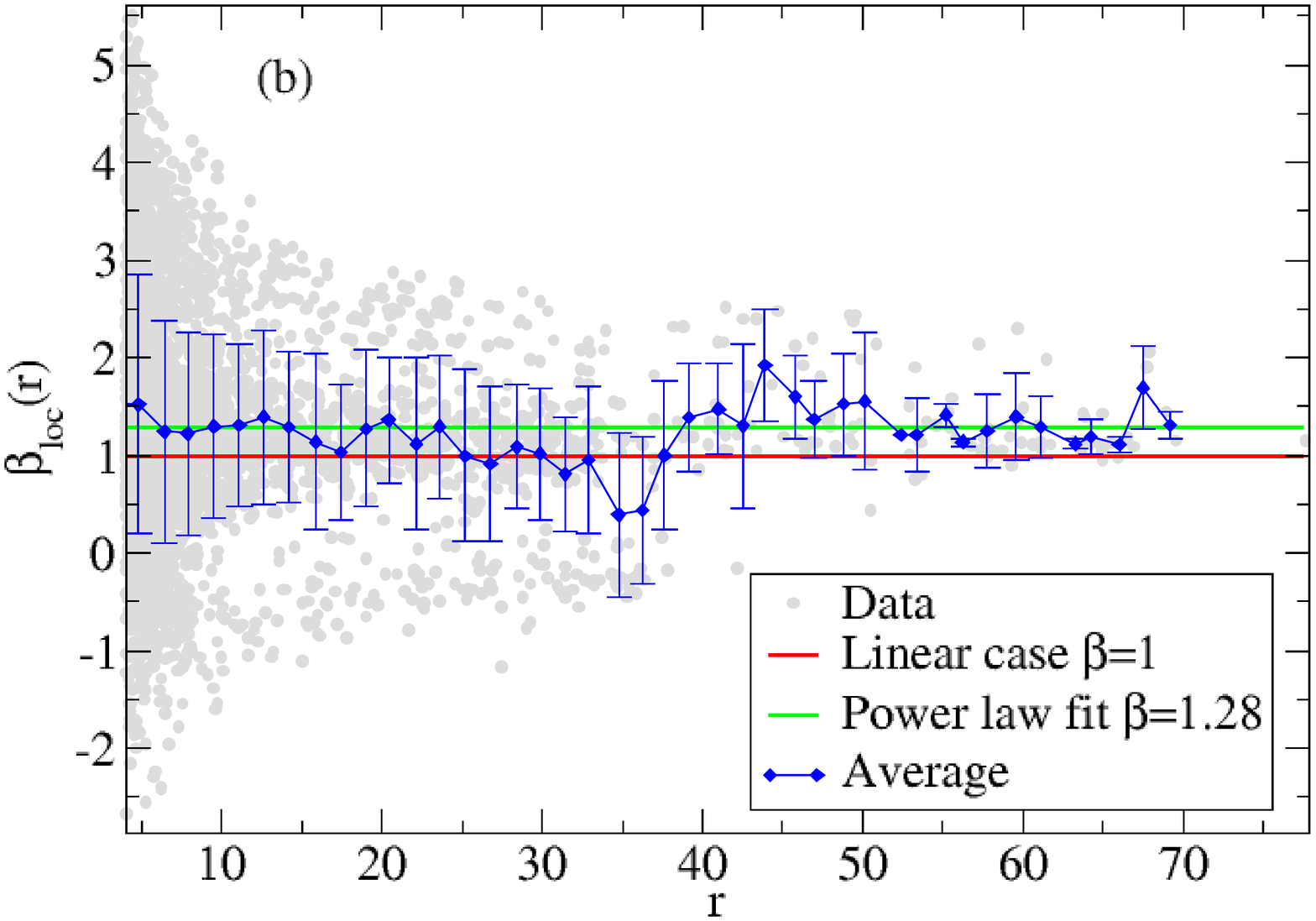}\\
  \includegraphics[width=0.9\linewidth]{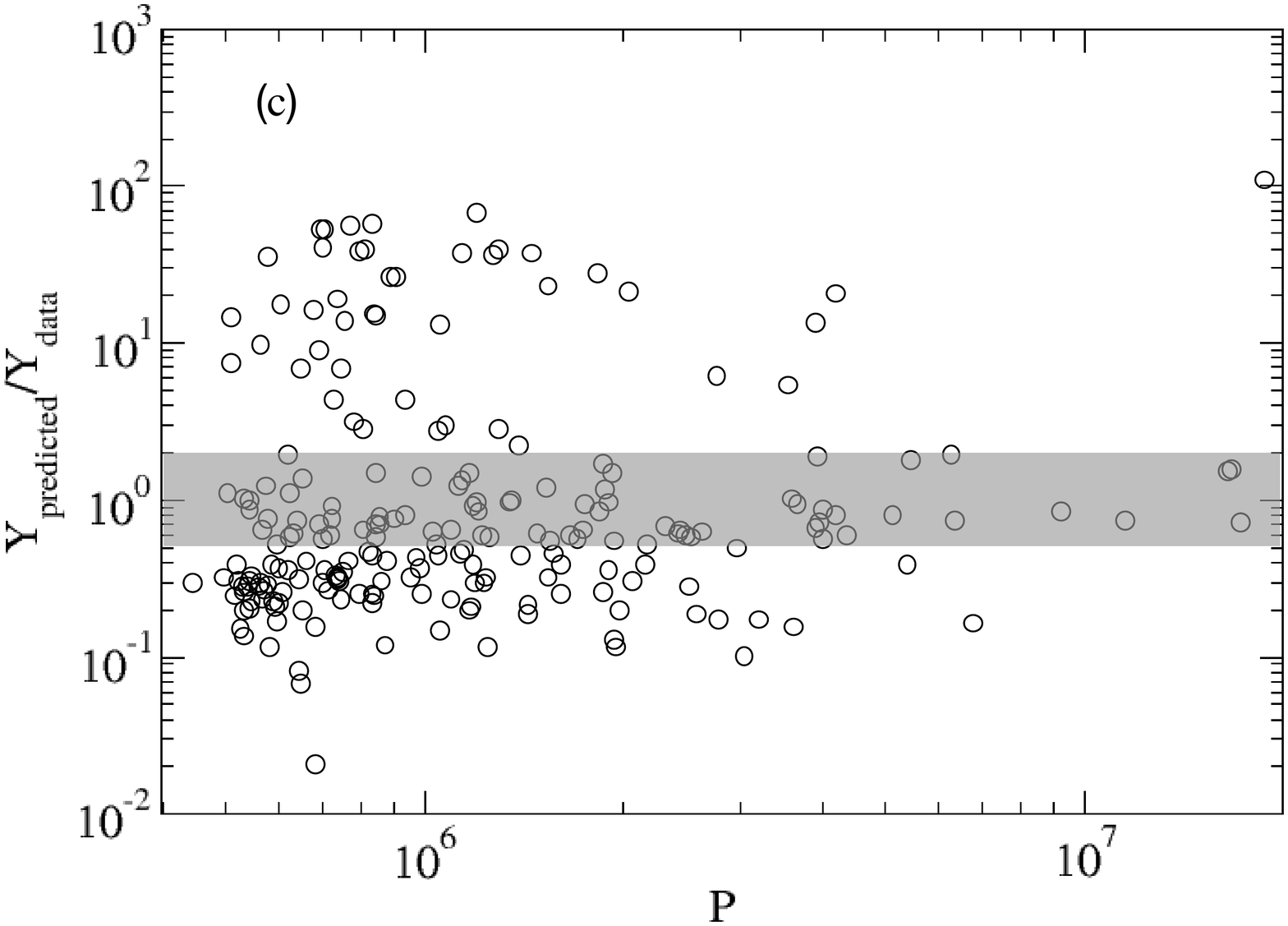}
  \caption{(a) Number of patents vs. population for cities in OECD countries \cite{Leitao:2016}. We show here the linear fit $aP$ with $a\approx 1.9 10^{-4}$ ($r^2=0.80$), and the power law fit $aP^\beta$ where $a\approx 1.23 10^{-6}$ and $\beta\approx 1.28$ ($r^2=0.53$). (b) Tomography plot for the number of patents in cities belonging to OECD countries. (c) Ratio of the predicted value over the real value. The grey area represents ratios that are in the range $[0.5,2]$.}
\label{fig:oecdpatents}
\end{figure}

In the case of patents in OECD cities, the authors of \cite{Leitao:2016} found a linear behavior. The plot of this number versus the population of cities is shown in Fig.~\ref{fig:oecdpatents}a.
We observe that there are large fluctuations and that both the linear and the nonlinear fit are consistent with the data (this is obviously due to the noise and the small number of available decades over which we can fit the data). The $r^2$ value for the linear fit is better and in agreement with the results of \cite{Leitao:2016} suggesting that the data follow a linear behavior. However, if we plot the local exponent (see Fig.~\ref{fig:oecdpatents}b), it seems that the superlinear behavior with $\hat{\beta}=1.28$ has a possible relevance to the data. The effective exponent $\beta_{\mathrm{eff}}=1.43$ is consistent with this superlinear behavior, but the fraction of cities with correct prediction is however small and about $37\%$ (see Fig.~\ref{fig:oecdpatents}c). At this stage, our analysis suggests that the behavior of OECD patents is neither linear nor superlinear, and probably not well represented by a simple scaling form. This might be due to the fact that we mix here different countries, with different economies, prohibiting a simple description in terms of a simple scaling function characterized by a single exponent.

\medskip
\paragraph{A note on scaling for patents.}

The number of patents is an important indicator for the productivity and innovation in cities and its study is therefore of great importance for understanding cities and what could be the critical factors for innovation \cite{Bettencourt:2007b}. We saw in previous sections that our analysis for patents in the UK shows a linear/slighly sublinear behavior and for OECD countries that the scaling form could be more complex than a single power law form.

\begin{figure}[ht!]
  \includegraphics[width=0.9\linewidth]{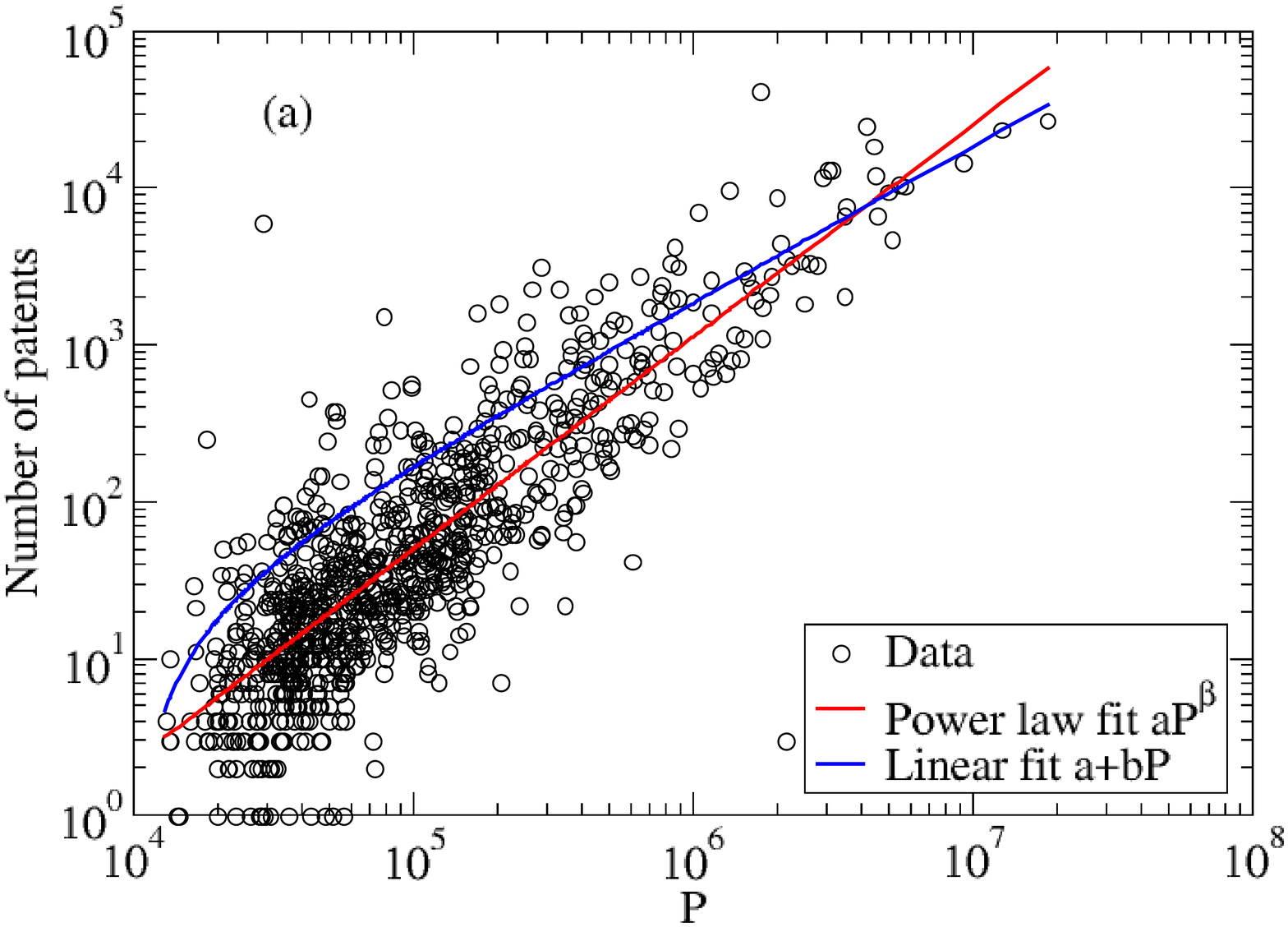}
    \includegraphics[width=0.9\linewidth]{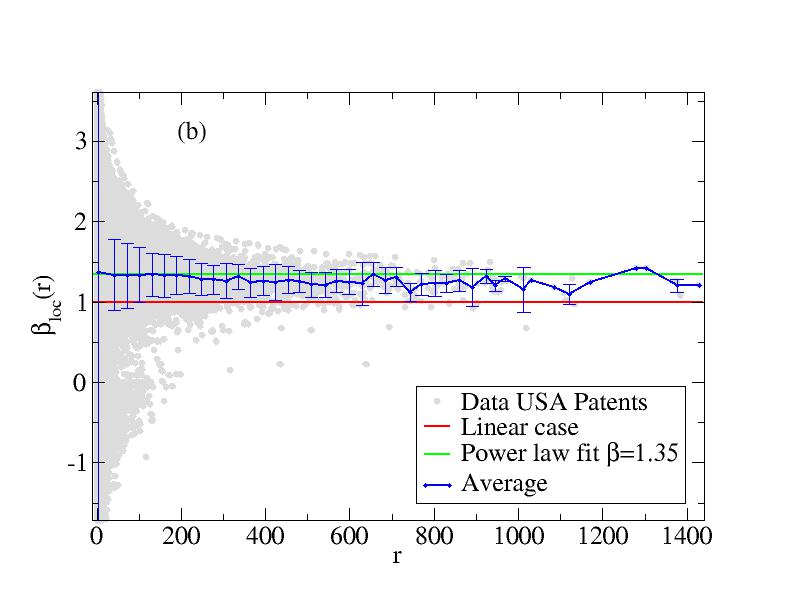}
  \caption{(a) Number of patents versus population for US cities (year 2005) \cite{Bettencourt:2010}. We show here both the power law fit ($\hat{\beta}=1.35\;(r^2=0.85)$) and the linear fit of the form $a+bP$ ($a=-19.4$ and $b=0.0018$, $r^2=0.85$). (b) Tomography plot for this quantity.}
\label{fig:usapatents}
\end{figure}

The case of US patents was not considered in \cite{Leitao:2016} but was studied in particular in \cite{Bettencourt:2010}. The power law fit for the 2005 US data (see \cite{Bettencourt:2010} for a detailed description of the dataset) gives $\hat{\beta}=1.35\;(r^2=0.85)$. However, a linear fit of the form $a+bP$ with $a=-19.4$ and $b=0.0018$ ($r^2=0.85$) is also consistent with data (Fig.~\ref{fig:usapatents}a). This last fit points to the possible existence of a threshold effect with a value $P_c\simeq 10,500$, an effect that might have some economical explanation. The tomography plot for this case is shown in Fig.~\ref{fig:usapatents}b and seems to confirm the superlinear behavior with a convergence of $\beta_{\mathrm{loc}}$ towards $\hat{\beta}$, in agreement with results discussed in \cite{Bettencourt:2007b}. The effective exponent is $\beta_{\mathrm{eff}}=1.19\pm 0.24$ and $f(1/2)=55\%$ confirming this superlinearity.

The UK, OECD countries and the US therefore display very different behavior for the scaling of the  number of patents and we summarize the results in the table \ref{table:patents}.
  \begin{table}[ht!]
    \caption{Results for patents for different regions: UK, USA, and OECD countries. For each case, we give the fitting exponent $\hat{\beta}$ (and the corresponding $r^2$ value), the effective exponent $\beta_{\mathrm{eff}}$ together with the value $f(1/2)$, and the conclusions of our analysis.}
\centering
\begin{tabular}{l c c c} 
\hline 
Region & Fit $\hat{\beta}$($r^2$) & $\beta_{\mathrm{eff}}$ ($f(1/2)$) & Conclusion\\ 
\hline 
  {\textit{UK}} & $1.06\;(0.88)$ & $0.96$  ($62\%$) & Linear\\
    &  &  & Slightly sublinear\\
\hline 
  \textit{USA} & $1.35\;(0.85)$ & 1.19 ($55\%$) & Superlinear\\
  \hline 
  \textit{OECD} & $1.285\;(0.53)$ & $1.43$  ($37\%$) & Superlinear\\ 
   &  &  & Other scaling form ?\\ 
\hline 
\end{tabular}
\label{table:patents} 
\end{table}

A possible reason for these different behaviors is that the level of aggregation is not the same for these three cases: The OECD is a collection of very different countries, the US is composed of states with various level of activity, and the UK is a much smaller set of countries. A study focused on the scaling of this quantity across different countries, at various level of aggregation might reveal more information and is certainly an interesting direction for future research.


\subsection{Inconclusive cases}

For some of the datasets studied in \cite{Leitao:2016}, standard tools could not lead to a clear conclusion about whether the scaling is linear or not. There are mainly two reasons for this. The first one is that $\beta$ can be larger or smaller than one depending on the assumption used for describing the fluctuations. The second reason is that for some cases the best model for fluctuations improves only marginally the statistics compared to the linear fit.  The concerned datasets of \cite{Leitao:2016} are the following. For Europe the cinema capacity and usage (reason 2), and the number of theatres (for the first reason), and in Brazil the number of deaths caused by external causes (second reason). These cases therefore represent interesting playgrounds for testing other methods. We will use the tools developed in this paper and will show that our method can bring some new conclusions or a new perspective such as the existence of a threshold for example.

\subsubsection{Cinema capacity and usage (Europe)}

We start with the cinema capacity (total number of seats) in European cities. The naive fit gives a linear behavior with $\hat{\beta}=0.99\;(r^2=0.71)$.
\begin{figure}[ht!]
  \includegraphics[width=0.9\linewidth]{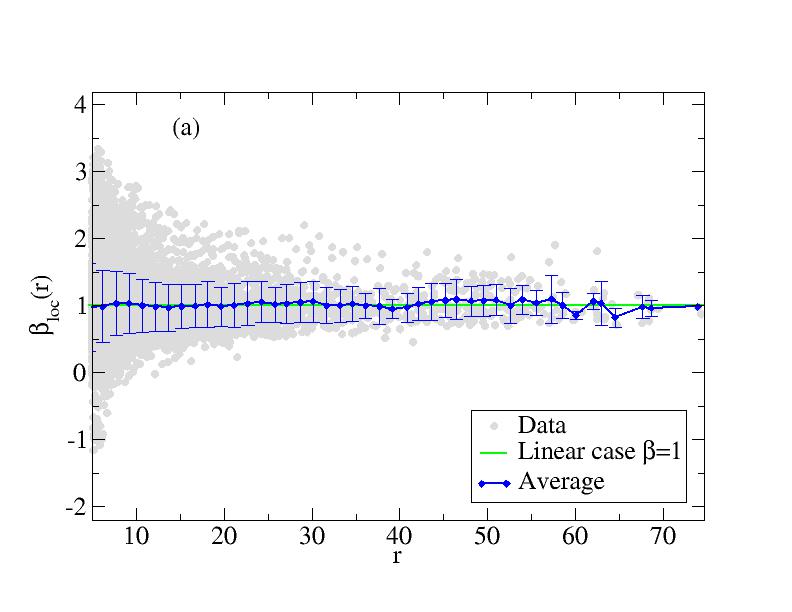}
  \includegraphics[width=0.9\linewidth]{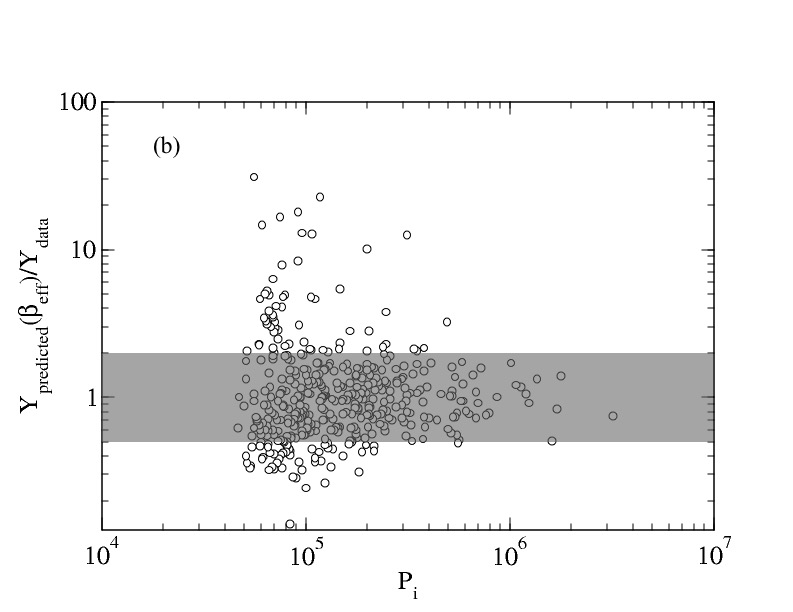}
  \caption{Cinema capacity in European cities. (a) Tomography plot and (b) ratio $Y_{predicted}/Y_{data}$ versus the population.  The gray area represent the fraction of cities with ratio in $[0.5,2.0]$ and which is about $74\%$ here..}
\label{fig:eurocinema}
\end{figure}
The tomography plot confirms this: there is a convergence of $\beta_{\mathrm{loc}}$ to $1$ (see Fig.~\ref{fig:eurocinema}a). The calculation of the effective exponent gives $\beta_{\mathrm{eff}}=0.98$ and for this value the fraction of cities with a prediction in $[0.5,2.0]$ is $f(1/2)=74\%$. All these results point in favor of a linear behavior. Even if the statistical evidence found in \cite{Leitao:2016} for this behavior seemed not to be large enough, we have here an objective $74\%$ of cities whose cinema capacity is correctly predicted using an exponent equal to $0.98$.

\begin{figure}[ht!]
  \includegraphics[width=0.9\linewidth]{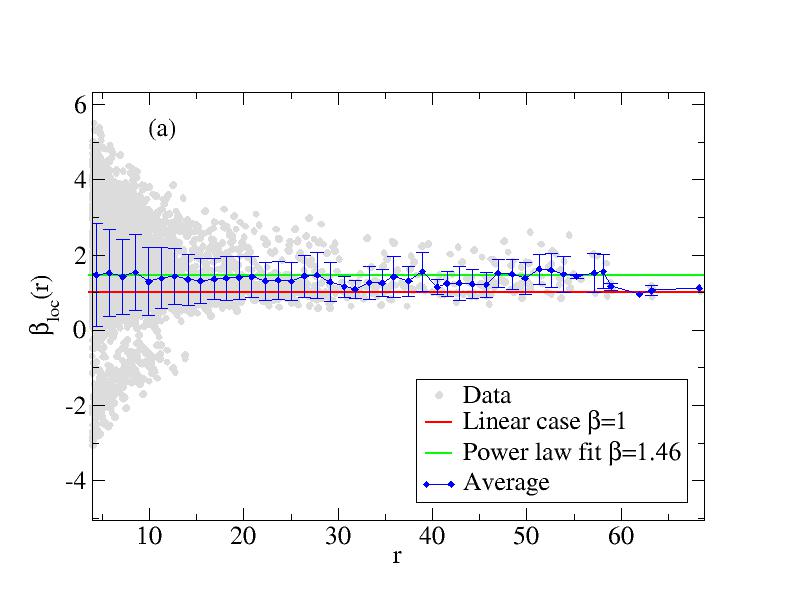}
  \includegraphics[width=0.9\linewidth]{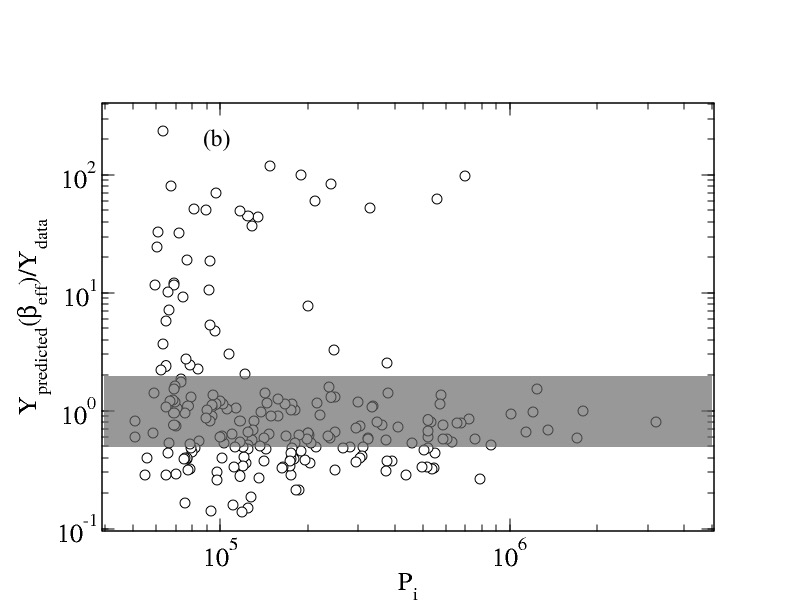}
  \caption{Cinema usage in European cities for the year 2011. (a) Tomography plot. (b) Ratio $Y_{predicted}/Y_{data}$ versus the population.  The gray area represent the fraction of cities with ratio in $[0.5,2.0]$ and which is here about $50\%$.}
\label{fig:eurocinemausage}
\end{figure}

In the case of cinema usage computed as the attendance in cinemas in the year 2011, the power law fit gives the exponent $\hat{\beta}=1.46\;(r^2=0.64)$ indicating a strongly nonlinear behavior. The tomography plot is shown in Fig.~\ref{fig:eurocinemausage}a, 
and shows that for most pairs of cities the local exponent is larger than $1$, except for very large ratios $r\gtrsim 60$. This suggests that there is a tendency towards a nonlinear behavior in agreement with the value of the effective exponent that we find $\beta_{\mathrm{eff}}=1.17$. For this value however, the ratio $Y_{predicted}/Y_{data}$ (shown in Fig.~\ref{fig:eurocinemausage}b)  indicates large fluctuations with only about $50\%$ of cities with a ratio $Y_{predicted}/Y_{data}$  in the range $[0.5,2.0]$. The other $50\%$ display a ratio either in $[0.1,0.5]$ or much larger, up to $10^2$ (a picture that is be confirmed by the slow increase of the function $f(1/\varepsilon,\varepsilon)$ with $\varepsilon$, not shown). In this respect, with such large fluctuations it is indeed a bit hard to conclude, although the superlinear behavior with $\beta_{\mathrm{eff}}=1.17$ accounts for half of the cities.

\subsubsection{Theaters in Europe}

\begin{figure}[ht!]
  \includegraphics[width=0.9\linewidth]{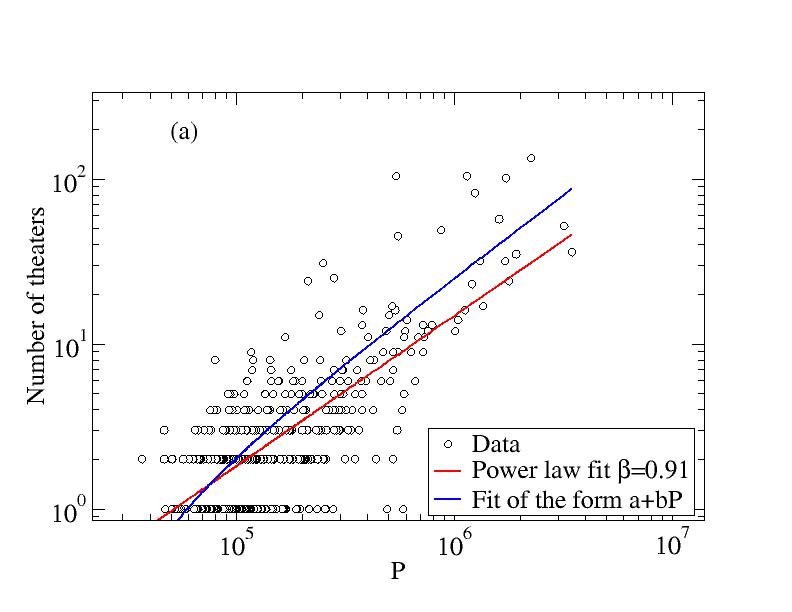}
  \includegraphics[width=0.9\linewidth]{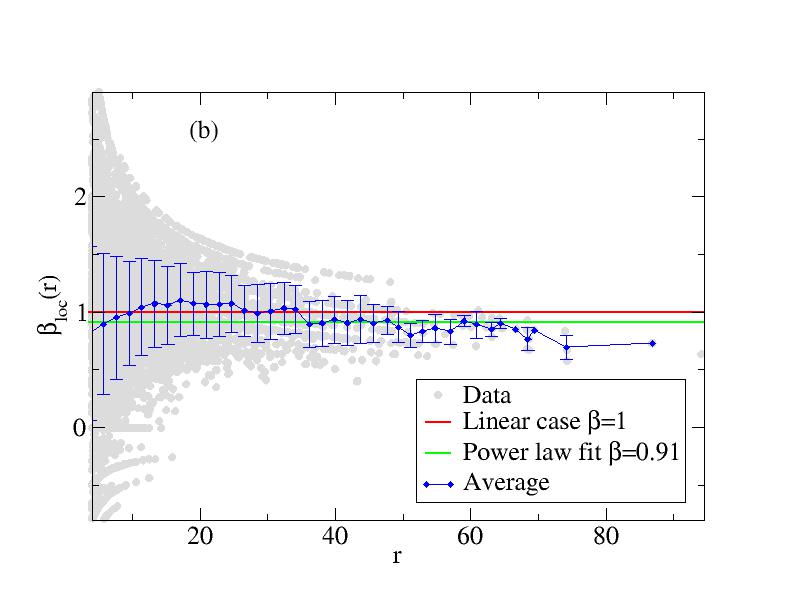}
  \caption{(a) Number of theaters in European cities versus their population.  The red line is the power law fit with exponent $\hat{\beta}=0.91$ ($r^2=0.74$). The green line is the fit of the form $a+bP$ with $a=-0.51$, and $b=2.5 10^{-5}$ ($r^2=0.68$). This fit implies a threshold value $P_c\approx 20,400$. (b) Tomography plot: local exponent versus population ratio $r$ for the number of theaters in European cities.}
\label{fig:eurotheater}
\end{figure}

This dataset contains the number of theaters in European cities (for the year 2011).
This case was classified as inconclusive in \cite{Leitao:2016} as the exponent value for $\beta$ could be either larger or smaller than one depending on the assumptions about the fluctuations. Despite large fluctuations, we can try a power law fit and the corresponding exponent is $\hat{\beta}=0.91\;(r^2=0.74)$ (Fig.~\ref{fig:eurotheater}a). The tomography plot (Fig.~\ref{fig:eurotheater}b) confirms indeed that this is a difficult case: for $r\lesssim 40$, the local exponent is around 1 while for larger values we observe local exponents smaller than 1 and even smaller than $\hat{\beta}$. There is therefore no clear convergence towards the fitting value and this might explain why this case, despite a relatively clear sublinearity, was considered as inconclusive in \cite{Leitao:2016}. This forces us to reconsider the validity of the power law fit, knowing that we have essentially one decade of variations which is far from being enough for a good fit. We note here that a fit of the form $a+bP$ (or obviously a more complex one of the form $a+bP^{\beta}$) with $a=-0.51$, and $b=2.5 10^{-5}$ ($r^2=0.68$) is also consistent with data.  This last fit implies a threshold value $P_c\approx 20,400$ above which the number of theatres is non-zero. We note that a threshold effect is here somehow expected: indeed only for cities large enough we observe the appearance of theaters. If we however try to compute the effective exponent we obtain $\beta_{\mathrm{eff}}=0.95$ and the corresponding fraction is $f(1/2)=71\%$, suggesting here a slightly sublinear behavior. This effective exponent together with the tomography plot therefore suggest a slight sublinear behavior, but we cannot exclude the possibility of a threshold effect (which are not mutually exclusive properties).

\subsubsection{Brazil: death by external causes}

This database is provided by Brazil’s Health Ministry for the year 2010 and gives the number of deaths by external causes. In this case too, the authors of \cite{Leitao:2016} found that there were not enough statistical evidences in order to conclude. 
\begin{figure}[h!]
 \includegraphics[width=0.9\linewidth]{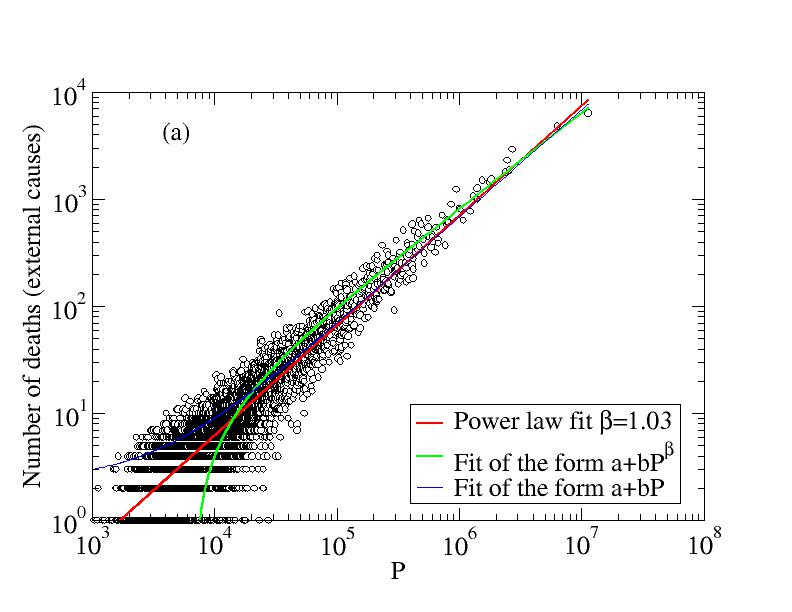}
  \includegraphics[width=0.9\linewidth]{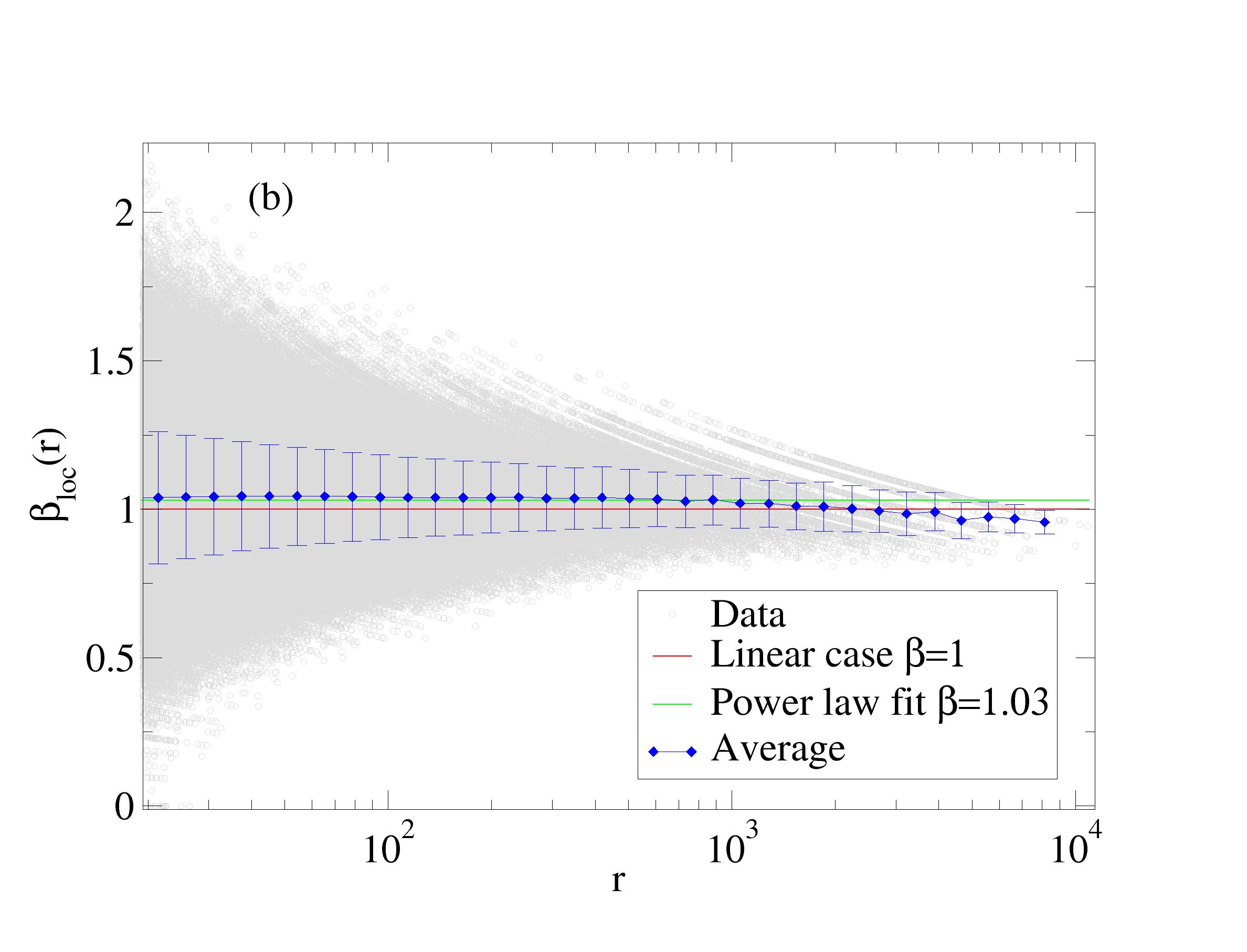}
  \caption{Death by external causes in Brazil (year 2010). (a) Number of deaths versus the population. We show here various fits including the power law fit $aP^\beta$ with $\beta=1.03$ ($r^2=0.91$), and fits of the form $a+bP$ (with $a=2.31$, $b=0.00068$, $r^2=0.97$) or $a+bP^\beta$ (with $a=-9.40$, $b=0.0035$, $\beta=0.90$, $r^2=0.98$). (b) Tomography plot showing a convergence towards an exponent equal to $1$.}
\label{fig:brazilexternal}
\end{figure}
We show this number versus the population in Fig.~\ref{fig:brazilexternal}a for various fits. The power law is not too bad and predicts a linear behavior $\hat{\beta}=1.03$. The forms $a+bP$ and $a+bP^\beta$ however don't produce consistent results about the existence of a threshold effect: for the linear fit there is no threshold while for the second fit (of the form $a+bP^{\beta'}$) there would be a small threshold value $P_c=6,500$ (and $\beta'=0.90$). It is therefore hard to conclude at this stage but the tomography plot shown in Fig.~\ref{fig:brazilexternal}b is rather clear and points to a linear behavior: the local exponent converges quickly towards 1 and its average is equal to one for all values of $r$ (within error bars). The effective exponent computed for this case is $\beta_{\mathrm{eff}}=0.99$ and the fraction of cities whose number of deaths is correctly predicted with this value is about $82\%$. Despite the difficulties with fitting the original data we have here an interesting case where the local exponent analysis is clear and all evidences point to a linear scaling.

\section{Discussion}

We summarize all our results in the table 1 where we compare them to the conclusions of \cite{Leitao:2016}.
\begin{table*}[ht!]
\caption{Exponent $\hat{\beta}$ obtained by fitting the data with the corresponding $r^2$ value and if needed the value of a possible threshold value $P_c$; effective exponent obtained by minimizing the error in predicting quantities for cities; fraction $f(1/2)$ of cities for which the ratio predicted value/actual value is in the range $[0.5,2]$; conclusions of the statistical analysis of \cite{Leitao:2016} and our conclusions.}
\centering
\begin{tabular}{l c c c c c} 
\hline\hline 
Data & Fit $\hat{\beta}$($r^2$) & $\beta_{\mathrm{eff}}$ & $f(1/2)$ & Conclusions of \cite{Leitao:2016} & Our conclusions\\ [0.5ex] 
\hline 
{\textit{UK}} & & & & &\\
\;\; Income & $1.01\;(0.99)$ & $0.966\pm 0.035$           & $100\%$ & Linear & Linear/slightly sublinear\\
\;\; Railroads & $0.50\;(0.76)$, $P_c=30,084$        &  $0.12\pm 0.17$           & $85\%$ & Linear & Threshold effect\\
\;\; Patents & $1.06\;(0.88)$ & $0.96\pm 0.13$                & $62\%$ & Linear & Slightly sublinear\\
\hline 
\textit{USA} & & & & &\\
\;\; GDP    & $1.11\;(0.98)$ & $1.13\pm 0.07$  & $98\%$ & Superlinear & Superlinear\\
\;\; Roads & $0.85\;(0.91)$ & $0.80\pm 0.1$    & $91\%$ & Sublinear & Sublinear\\ [1ex] 
\hline 
\textit{Europe} & & & & &\\
\;\; Cinema (capacity)    & $0.99\;(0.71)$ & $0.98\pm 0.29$              & $74\%$ & Inconclusive & Linear\\
\;\; Cinema (usage)       & $1.46\;(0.64)$ & $1.17\pm 0.55$              & $50\%$ & Inconclusive  & Superlinear (large fluctuations) \\
\;\; Museum (usage)     & $1.42\;(0.69)$ & $1.64\pm 0.47$              & $45\%$ & Superlinear & Superlinear (large fluctuations)\\
\;\; Theatres                 & $0.91\;(0.74)$ $P_c=20,400$  & $0.95\pm 0.45$     & $71\%$ & Inconclusive & Slightly sublinear/Threshold effect\\
\;\; Libraries                & $0.80\;(0.59)$ & $0.17\pm 0.32$ & $55\%$ & Sublinear &Fluctuations too large\\ [1ex] 
\hline 
\textit{OECD} & & & & &\\
\;\; GDP    & $1.12\;(0.91)$    & $1.06\pm 0.16$  & $90\%$ & Superlinear & Superlinear\\
\;\; Patents & $1.285\;(0.53)$ & $1.43\pm 0.71$  & $37\%$ & Linear  & Superlinear/no simple form\\ [1ex] 
\hline 
\textit{Brazil} & & & & &\\
\;\; GDP       & $1.04\;(0.86)$       & $1.21\pm 0.11$  & $63\%$ & Superlinear   & Superlinear\\
\;\; AIDS      & $0.74\;(0.81)$, $P_c=10,080$       & $1.03\pm 0.13$  & $67\%$ & Sublinear      & Linear/Threshold effect\\
\;\; External & $1.03\;(0.91)$ & $0.99\pm 0.08$ & $82\%$ & Inconclusive & Linear\\ [1ex] 
\hline 
\end{tabular}
\label{table:exp} 
\end{table*}
We developed here simple tools for analysing data that could help for understanding their scaling behavior. Although these tools do not replace the standard statistical analysis, they enable a more practical view of the system's behavior: if we had to use the scaling form for making predictions what would be the most reliable exponent ? One advantage of this approach is that the answer to this question does not depend on some assumptions, such as the nature of the noise for example. In cases where noise is small and the number of available decades is large, our analysis simply confirms standard tools such as fitting methods. It is in more complex cases where it is difficult to decide which model describes the best the data that our method could be of some help. The analysis of the local exponent gives a precise picture of how different systems of different sizes are related to each other. In some cases it allows to conclude with more confidence about the nonlinear or linear behavior, but in other cases it also signals the failure of a simple scaling. This failure could happen due to a threshold effect for example, but more generally, we could expect that the system is described by a more complex function with more than one exponent for example. It would be interesting to apply this method at various level of aggregation for a given quantity, but also to test the temporal evolution of a system as it might reveal some information about its dynamics.

We could summarize this analysis by proposing the following set of necessary conditions in order to
trust the fitting value $\hat{\beta}$:
\begin{itemize}
\item{} (i) We need the convergence of
  $\beta_{\mathrm{loc}}$ towards $\hat{\beta}$.
  \item{} (ii) The value of the effective
    exponent $\beta_{\mathrm{eff}}$ should be consistent with $\hat{\beta}$. In general the value $\beta_{\mathrm{eff}}$ should be preferred over $\hat{\beta}$, in particular if the value $f(1/2)$ is large (see (iii)).
    \item{} (iii) The value of $f(1/2)$ should be at least $50\%$. This value could of course be debated but at least we should observe a rapid increase of $f(\varepsilon,1/\varepsilon)$ with decreasing $\varepsilon$.
  \end{itemize}
  If these conditions are not satisfied, we can safely reject the value obtained $\hat{\beta}$ obtained by the power law fit. In this case, it suggests for example that either fluctuations are too large or that the simple power law scaling form is not valid. 
  
The discussion was done here on urban data but this method could obviously be applied to any system that displays scaling. In addition, we could probably envision other measures here, but we believe that this sort of bootstraping could help to better understand the scaling in complex systems, to circumvent lengthy and often non convergent debates about the quality of a fit.

\medskip

{\bf Acknowledgments}. I thank Luis Bettencourt,  Jos\'e Lobo, Scott Ortman, Michael Smith for having organized at the Santa Fe Institute a workshop entitled `Integrating Views on Urban Scaling: Foundations, Criticisms, and Extensions', on May 15-17, 2019. During this workshop I had the chance to discuss with Elsa Arcaute, Luis Bettencourt, Markus Hamilton, Jos\'e Lobo, Scott Ortman, Celine Rozenblat, Diego Rybski, Michael Smith, Deborah Strumsky, Geoffrey West, and David White and I thank them for stimulating and challenging debates. Finally, I thank Elsa Arcaute, Luis Bettencourt, Diego Rybski, and Deborah Strumsky for having shared their data with me.



\begin{thebibliography}{99}

\bibitem{Kleiber:1932}
 Kleiber M. Body size and metabolism. Hilgardia, Vol. 6, no 11, 1932.

\bibitem{Kleiber:1947}
Kleiber M. Body size and metabolic rate. Physiological reviews. 1947 Oct 1;27(4):511-41.

\bibitem{degennes:1979}
De Gennes PG, Gennes PG. Scaling concepts in polymer physics. Cornell university press; 1979.

\bibitem{Goldenfeld:1992}
Goldenfeld N. Lectures on phase transitions and the renormalization group. Perseus Books Publishing, 1992. 

\bibitem{Barenblatt:2003}
Barenblatt GI. Scaling. Cambridge University Press; 2003.

\bibitem{West:1997}
West GB, Brown JH, Enquist BJ. A general model for the origin of allometric scaling laws in biology. Science. 1997 Apr 4;276(5309):122-6.

\bibitem{Pumain:2004}
Pumain, Denise. Scaling laws and urban systems. Working paper SFI. (2004).

\bibitem{Bettencourt:2007a}
Bettencourt LM, Lobo J, Helbing D, Kühnert C, West GB. Growth, innovation, scaling, and the pace of life in cities. Proc. Natl. Acad. Sci. (USA) 104(17):7301-6 (2007).

\bibitem{Batty:2013}
Batty M. The new science of cities. MIT press; 2013.

\bibitem{Barthelemy:2016}
Barthelemy M. The structure and dynamics of cities. 
Cambridge University Press; 2016 Nov 24.

\bibitem{Barthelemy:2019}
Barthelemy M. The statistical physics of cities. 
Nature Reviews Physics. 2019 May 3:1.

\bibitem{Samaniego:2008}
Samaniego H, Moses ME. Cities as organisms: Allometric scaling of urban road networks. Journal of Transport and Land use. 2008 Jul 1;1(1):21-39.

\bibitem{Fuller:2009}
Fuller RA, Gaston KJ. The scaling of green space coverage in european cities. Biology letters, vol. 5, no. 3,
pp. 352–355, 2009.

\bibitem{Kuhnert:2006}
Kuhnert C, Helbing D, West GB. Scaling laws in
urban supply networks. Physica A: Statistical Mechanics
and its Applications, vol. 363, no. 1, pp. 96–103, 2006.

\bibitem{Fragkias:2013}
Fragkias M, Lobo J, Strumsky D, Seto KC. Does size matter? Scaling of CO2 emissions and US urban areas. PLoS One. 2013 Jun 4;8(6):e64727.

\bibitem{Rybski:2013}
Rybski D, Sterzel T, Reusser DE, Winz AL, Fichtner C, Kropp JP (2013) Cities as nuclei of stability? arXiv:1304.4406

\bibitem{Oliveira:2014}
 Oliveira EA, Andrade JS, Makse HA (2014). Large cities are less green. Scientific Reports 4:4235.

\bibitem{Oliveira:2019}
Ribeiro HV, Rybski D, Kropp JP (2019) Effects of changing population or density on urban carbon dioxide emissions. Nature communications 10.

\bibitem{Rybski:2009}
Rybski D, Buldyrev SV, Havlin S, Liljeros F, Makse HA. Scaling laws of human interaction activity. Proceedings of the National Academy of Sciences. 2009 Aug 4;106(31):12640-5.

\bibitem{Bettencourt:2010}
Bettencourt LM, Lobo J, Strumsky D, West GB. Urban scaling and its deviations: Revealing the structure of wealth, innovation and crime across cities. PloS one. 2010 Nov 10;5(11):e13541.

\bibitem{Khiali:2019}
Khiali-Miab A, van Strien MJ, Axhausen
KW, Grêt-Regamey A (2019) Combining urban
scaling and polycentricity to explain socioeconomic status of urban regions. PLoS ONE 14
(6): e0218022.

\bibitem{Alves:2013}
Alves LGA, Ribeiro HV, Lenzi EK, Mendes RS. 2013
Distance to the scaling law: a useful approach for
unveiling relationships between crime and urban
metrics.PLoS ONE 8, e69580. (doi:10.1371/journal.
pone.0069580)

\bibitem{Lobo:2013}
Lobo J, Bettencourt LM, Strumsky D, West GB. Urban scaling and the production function for cities. PLoS One. 2013 Mar 27;8(3):e58407.

\bibitem{Bettencourt:2007b}
Bettencourt LM, Lobo J, Strumsky D. Invention in the city: Increasing returns to patenting as a scaling function of metropolitan size. Research policy. 2007 Feb 1;36(1):107-20.

\bibitem{Nomaler:2014}
Nomaler O, Frenken K, Heimeriks G. 2014 On scaling
of scientific knowledge production in U.S.
metropolitan areas.PLoS ONE 9, e110805.
(doi:10.1371/journal.pone.0110805)

\bibitem{Strano:2016}
  Strano E, Sood V. Rich and poor cities in europe. An urban scaling
  approach to mapping the european economic transition. PloS one, vol. 11, no. 8, p. e0159465,
2016.

\bibitem{Caminha:2017}
Caminha C, Furtado V, Pequeno TH, Ponte C, Melo HP, Oliveira EA, Andrade Jr JS (2017) Human mobility in large cities as a proxy for crime. PloS one, 12:e0171609.



\bibitem{Pumain:2006}
Pumain D, Paulus F, Vacchiani-Marcuzzo C, Lobo J. An evolutionary theory for interpreting urban scaling laws. Cybergeo: European Journal of Geography. 2006 Jul 5.

\bibitem{Bettencourt:2013}
Bettencourt LM. The origins of scaling in cities. science. 
2013 Jun 21;340(6139):1438-41.

\bibitem{Bettencourt:2013b}
Bettencourt LMA, Lobo J, Youn H. 2013 The
hypothesis of urban scaling: formalization,
implications and challenges. arXiv:1301.5919.

\bibitem{Louf:2013}
Louf R, Barthelemy M. Modeling the polycentric transition of cities. Physical review letters. 2013 Nov 6;111(19):198702.

\bibitem{Louf:2014}
Louf R, Barthelemy M. How congestion shapes cities: from mobility patterns to scaling. Scientific reports. 2014 Jul 3;4:5561.

\bibitem{Verbavatz:2019}
Verbavatz V, Barthelemy M. Critical factors for mitigating car traffic in cities. PLoS ONE 14(7): e0219559 (2019)

\bibitem{Molinero:2019}
  Molinero C, Thurner S. How the geometry of cities explains urban scaling laws and determines their exponents. arXiv:1908.0747.

\bibitem{bea}
  Bureau of Economic Analysis: \url{https://apps.bea.gov/iTable/iTable.cfm?isuri=1&reqid=70&step=1#isuri=1&reqid=70&step=1} (accessed 20.08.2019).

\bibitem{Shalizi:2011}
Shalizi CR. Scaling and hierarchy in urban economies. arXiv preprint arXiv:1102.4101, 2011.

\bibitem{Leitao:2016}
Leitao JC, Miotto JM, Gerlach M, Altmann EG. Is this scaling nonlinear? Royal Society open science. 2016 Jul 1;3(7):150649.

\bibitem{Arcaute:2015}
Arcaute E, Hatna E, Ferguson P, Youn H, Johansson A, Batty M. Constructing cities, deconstructing scaling laws. Journal of The Royal Society Interface. 2015 Jan 6;12(102):20140745.

\bibitem{Cottineau:2017}
Cottineau C, Hatna E, Arcaute E, Batty M. 
Diverse cities or the systematic paradox of urban 
scaling laws. Computers, environment and urban systems. 2017 May 1;63:80-94.

\bibitem{Louf:2014}
Louf R, Barthelemy M. Scaling: lost in the smog. Environment and Planning B: Planning and Design. 2014 Oct;41(5):767-9.

\bibitem{Depersin:2018}
Depersin J, Barthelemy M. From global scaling to the dynamics of individual cities. Proceedings of the National Academy of Sciences. 2018 Mar 6;115(10):2317-22.

\bibitem{Ockham}
 Ockham W. Philosophical writings. Hackett Publising 1990.

\bibitem{Bettencourt:2019}
  Bettencourt L. Complex Networks and Fundamental Urban Processes.
  Mansueto Institute for Urban Innovation Research Paper. 2019 Jun 11(9).
  
\bibitem{Coniglio:1989}
Coniglio A, Zannetti M. Multiscaling in Growth Kinetics. Europhys. Lett., 10 (6), pp. 575-580 (1989). 


\end{thebibliography}


\end{document}